# Entropy analyses of spatiotemporal synchronizations in brain signals from patients with focal epilepsies

Çağlar Tuncay
caglart@metu.edu.tr

**Abstract**: The electroencephalographic (EEG) data intracerebrally recorded from 20 epileptic humans with different brain origins of focal epilepsies or types of seizures, ages and sexes are investigated (nearly 700 million data). Multi channel univariate amplitude analyses are performed and it is shown that time dependent Shannon entropies can be used to predict focal epileptic seizure onsets in different epileptogenic brain zones of different patients. Formations or time evolutions of the synchronizations in the brain signals from epileptogenic or non epileptogenic areas of the patients in ictal interval or inter-ictal interval are further investigated employing spatial or temporal differences of the entropies.

**Introduction**: Automatic (computer based) detection or prediction of epileptic seizures is a challenging subject in quantified electroencephalography [1-13] where the underlying dynamics for the recordings are usually not completely known. Thus, it is difficult to select a priori most suitable method for an analysis [14].

Fortunately, the ictal state is characterized by occurrence of synchronous oscillations and two scenarios of how a spontaneous seizure could evolve are suggested in [15-16]: If a seizure is caused by a sudden and abrupt transition, then it would not be preceded by detectable dynamical changes in EEG data. Such a scenario is suggested for the initiation of seizures in primary generalized epilepsy. On the other hand, if the transition occurs gradually then it could be detected, at least in principle. This type of transition is proposed to be more likely in focal epilepsies. In this case, the seizure begins in a restricted brain region and either remains localized or spreads to the adjacent cortex.

These hypotheses are considered and tested by various methods in several papers. For example in [17 and 18], the interdependence between intracranial EEG electrodes in human subjects with epilepsy is investigated. It is reported in both publications that there is strong evidence for non-linear interdependence between the focus of epileptic activity and other brain regions and this interdependence can be detected for up to 30 seconds prior to a seizure. Different researchers [19] studied single electrode scalp EEG from 60 healthy humans and found that non-linear structure is only present weakly and infrequently. More recently, the same hypotheses were tested by comparing 30 measures in terms of their ability to distinguish between the inter-ictal period and the pre-seizure period [20 and 21]. These publications provide statistically significant evidence for the existence of a pre-ictal state which is supported by numerous clinical evidence involving an increase in cerebral blood flow [22-23], oxygen availability [24] and blood-oxygen-level-dependent signal [25] as well as changes in heart rate [26-28] before seizure onsets. Thus, the interdependence between several brain sites and the existence of pre-ictal state may be important for computer based investigations of focal epilepsy.

The above mentioned approaches based on amplitude, time or frequency domain measurements [21, 29-31] have a shared property in that they all try to figure out whether there is any common information between the EEG time series as an indication of their relationship. Then, direct investigations by means of information-theoretic approaches can be favored [29]. Following this idea in this work, multi channel univariate time dependent entropy analyses are performed on various electroencephalograms intracerebrally recorded from humans with focal epilepsies [32]. The basic aim is to predict and detect numerous seizure onsets covered in the investigated data. In the meantime, several properties in the EEG data will be disclosed and used to distinguish between the inter-ictal period and the pre-seizure period [20-21]. It is clear that a method capable of predicting the occurrence of seizures from the electroencephalograms of epilepsy patients would open new therapeutic possibilities.

*Time dependent entropy*: In this study, time dependent interrelated multi channel EEG activities of numerous humans are investigated. The ability of an information measure (entropy) index [6, 33, 34], which is defined below, is determined to distinguish between pre-ictal, ictal or post-ictal periods as well as the inter-ictal interval. The multi channel univariates are cut into consecutive sections of convenient length of 10 seconds, and then entropy is calculated for the distribution of data in each of the windows which do not overlap [35, 36]. In other words, the analyses are performed in a time window of defined length and this time window is moved step wise to produce a continuously updated entropy spectrum for the recordings from different brain sites. Note that the model used is time invariant as stipulated in [35].

Entropy is known to be a quantity describing the amount of disorder in a system or a measure of uncertainty in the information content of a signal. Shannon entropy [37 and 38] is used to investigate various properties of integrated neuronal activity in this work, where the recorded voltages ($X_k(i)$) from the contact position (k=1-6) have integer values in micro Volt ($\mu V$) [32]. Let $Q(X_k(i))$ be the frequency of $X_k(i)$ in the data and be abbreviated as $Q_k(i)$. The (discrete) empirical distribution $P_k(i)$ is obtained by normalizing $Q_k(i)$;

$$P_k(i) = Q_k(i)/\sum_{i=1;N} Q_k(i) \quad , \tag{1}$$

where N is the number of samples (which is 2560 per window since the sampling rate is 256 Hertz (Hz) in [32] and the length of a time window is 10 seconds in this contribution). In what follows, ED refers to empirical distribution and an ED is said to be discrete uniform (or homogeneous) if $P(i) \propto 1/N$ or delta-like if data are distributed in a narrow peak around a voltage.

Then estimation of Shannon entropy ($S_k$) of a time window of a univariate recorded from the contact position (k) is

$$S_k = -\sum_{i=1;N} P_k(i)\log(P_k(i)) \quad . \tag{2}$$

In Eq. (2), the summation is over the states (i) which are accessible with probability $P_k(i) \leq 1$ and log is logarithm with any base [37 and 38] which is taken as the natural logarithm, here.

Note that $S_k$ is zero for delta-like ED. Moreover, $S_k$ attains its maximum value which is $\ln(N)$ [29] if the ED of a given set of N samples is homogeneous. Thus, an entropy measure index ($\chi_k$) may be defined as

$$\chi_k = S_k/\ln(N) \tag{3}$$

which is zero for delta-like ED or 1 for homogeneous ED and $0<\chi_k<1$ otherwise. (See, Appendix for further discussion of the subject.) One may refer to [20, 21, 39, 40] for several utilizations of the parameter in Eq. (3) with different aims.

In this work, the ED for each time window starting at the time ($\tau$-$\Delta\tau$) and ending at the time ($\tau$) with $\Delta\tau$=10 seconds are calculated and a continuously updated entropy index spectrum $\chi_k(\tau)$ is obtained for each univariate recorded from the contact position k=1-6.

The number of samples, N in Eq. (3), is the same for the windows and hence it is possible to select these time windows with small values of $\chi_k$ in the time profile ($\chi_k(\tau)$) when the EEG signal in the window is narrow (oscillations with small magnitudes about a voltage). On the other hand, the value of $\chi_k$ is close to 1 when the EEG signals are broad (oscillations with large magnitudes). Numerous further criteria may be applied to predict the characteristics of a sample distribution, if needed [29, 39, 41-45].

*Space and time differences of the entropy indices*: Due to the occurrence of synchronous EEG oscillations in the ictal state as discussed in the Introduction section, it is expected that the time dependent indices ($\chi_k(\tau)$) become similar at times close to or during a seizure. In other words, the values of the indices $\chi_k(\tau)$ and $\chi_{k'}(\tau)$ are synchronously close to each other when the brain sites for the contact positions (k) and (k') are coupled. Thus, the pairwise differences $D_{k,k'}(\tau)$ of $\chi_k(\tau)$ and $\chi_{k'}(\tau)$

$$D_{k,k'}(\tau) = \chi_k(\tau) - \chi_{k'}(\tau) \qquad (4)$$

may provide information about the strength (with small absolute value of $D_{k,k'}(\tau)$) and duration (number of the time windows with small absolute value of $D_{k,k'}(\tau)$) of a coupling between two brain sites. Note that $D_{k,k'}(\tau)$ may be considered as the approximate spatial derivatives of the index profiles for those contacts which are closely positioned.

The seizure time terms last for several seconds or minutes and they are clearly short with respect to several hours (h) for the registration periods of the data from a patient. Hence, a new parameter (z) is used to help the inspection of the similarities in the time profiles of $\chi_k(\tau)$:

$$z(\tau) = 1 / \sum_{k=1;6, k \neq k'} \sum_{k'=1;6} |D_{k,k'}(\tau)| \qquad (5)$$

where (|.|) stands for the absolute value. Obviously, $z(\tau)$ defined in Eq. (5) is big when the values of the entropy indices ($\chi_k(\tau)$) are close to each other. Thus, the predictive power of the marker depends on the strength of the couplings before or during onsets.

In this work, the first and second order successive time differences of the entropy indices, $\chi'_k(\tau)$ and $\chi''_k(\tau)$, respectively are also investigated with the aim of obtaining a better description of the properties of integrated neuronal activity in ictal and inter-ictal intervals, where

$$\chi'_k(\tau) = \chi_k(\tau) - \chi_k(\tau-\Delta\tau) \qquad (6)$$

and $\qquad \chi''_k(\tau) = \chi'_k(\tau) - \chi'_k(\tau-\Delta\tau) \quad . \qquad (7)$

In Eqs. (6) and (7), $\Delta\tau$ is equal to 10 seconds which is the length of a time window. These time rates of change provide additional information for the formations and time evolutions of the synchronized behaviors in entropy indices ($\chi_k(\tau)$). Thus, the attempts to understand the underlying mechanisms of the synchronizations in the EEG signals of the patients entail the analysis of the time series $\chi'_k(\tau)$ and $\chi''_k(\tau)$ as well as $\chi_k(\tau)$ and the spatial differences ($\chi_k(\tau)$- $\chi_{k'}(\tau)$) in the presented work. Obviously, comparison of the distributions of these quantities is further illuminating.

*Correlations*: Auto-correlation and cross-correlation functions are widely used in the literature to investigate the interdependence between physiological signals [29-31]. However, the values of the spatial differences of the electrodes ($D_{k,k'}(\tau)$) or their distributions may be used confidently with the same aim. For example, narrow and high peaks around zero in these distributions will imply strong correlations and long durations, respectively between the related individual recordings ($X_k$ and $X_k$') or vice versa. As a result, sample distributions of the values of $D_{k,k'}(\tau)$ for several k or k' are used to investigate the strengths and durations of the correlations in this contribution.

*Material*: Some brief information about all of the patients (PatNM) is given in Table I where the patients are grouped with respect to their origins of epilepsy. It is declared in [32] that "The EEG database contains invasive EEG recordings of 21 patients suffering from medically intractable focal epilepsy." Yet, the data from the Pat02 is missing there. Thus the available data are from 20 patients. Let us follow the lines of the declaration: "In order to obtain a high signal-to-noise ratio, fewer artifacts, and to record directly from focal areas, intracranial grid-, strip, and depth-electrodes were utilized. The EEG data were acquired using a Neurofile NT digital video EEG system with 128 channels, 256 Hz sampling rate, and a 16 bit analogue-to-digital converter. Notch or band pass filters have not been applied.

For each of the patients, there are data sets called 'ictal' and 'inter-ictal', the former containing files with epileptic seizures and at least 50 minute pre-ictal data. The latter containing approximately 24 h of EEG-recordings without seizure activity. At least 24 h of continuous interictal recordings are available for 13 patients. For the remaining patients interictal invasive EEG data consisting of less than 24 h were joined together, to end up with at least 24 h per patient. For each patient, the recordings of three focal and three extra-focal electrode contacts is available."

This pool of data has been downloaded by nearly hundred research groups from different countries [46] and treated with several aims [47-52].

The data recorded from epileptogenic zones will be designated with (k=1-3) and those from non epileptogenic ones with (k=4-6) for each patient, here. As a result, data recorded from each of the six contact positions (k) in the ictal interval or in the inter-ictal interval of each patient are cut into consecutive sections of length of 10 seconds. These time windows include 2560 data per individual recordings (k) with 256 Hz sampling rate. Then, $\chi_k(\tau)$ is calculated for the distribution of the data in each of the non overlapping windows. Hence, six time dependent trajectories of the entropy measure ($\chi_k(\tau)$) are obtained for the data recorded from each patient in ictal interval and another 6 trajectories for the data from the same patient in the inter-ictal interval. The results are compared for a characteristic behavior significant to pre-ictal, ictal or inter-ictal states. The seizure terms of each patient are indicated in [32] and the given numbers are utilized to select the seizure time windows in this work. The one-hour-data-blocks with consecutive integer numbers are assumed to be continuous due to the declaration [32].

Please note that numerous pin-like topographies are expected to occur temporarily in the index profiles for the collective behaviors where the topographies will be upward if the signals broaden in the meantime, or vice versa; namely, downward pin-like patterns will appear if narrow signals accompany the synchronized interdependencies.

The original results are presented in the following section. The last section is devoted to discussion and conclusion.

**Results**: In this section the continuously updated entropy index spectrum for each of the individual recordings from different brain sites ($\chi_k(\tau)$ with k=1-6) of 20 patients with different origins of epilepsy, seizure types, ages and genders are analyzed. The results are exemplified for six patients with different origins of epilepsy, types of seizures, ages and sexes where various characteristic behaviors are found in the data from pre-ictal, ictal or post ictal terms. The available [32] electrode types or electrode contacts of the considered six patients are shown in Figure 1. The theoretical results for $\chi_k(\tau)$ or $z(\tau)$ are plotted with a linear axis on the left and logarithmic axis on the right, respectively where the common horizontal axis shows the time ($\tau$) with the unit of 1 h, for 360 windows and the seizure terms are indicated by arrows in Figures 2-5, 7, 9, 11, 13, 15, 18, 21, 22, 25. The number of the significant digits is taken three after the decimal point in the results. Moreover, the bin sizes of the distributions displayed in Figures 6, 8, 10, 12, 14, 16, 17, 19, 20, 23 are the same, which is 0.001 and that in Figure 24 is 0.002. The parameter for k in the legends of the figures indicates the electrode number.

*CASE 1: Pat03 with Frontal Lobe Epilepsy*      The time spectra of the entropy indices ($\chi_k(\tau)$) with k=1-6 of the EEG data recorded from the Pat03 during the ictal interval and the inter-ictal interval are illustrated in Figs. 2 (a) and (b), respectively. It may be inspected that all of the (four out of four) indicated seizures [32] are detected by means of the marker $z(\tau)$ in Fig. 2 (a). Thus, all of the seizures of the patient are accompanied by strong simultaneous couplings (synchronous interdependency) between the brain sites for the contact positions.

Please note that the patterns of the trajectories of the entropy indices in the ictal interval and in the inter-ictal interval are different: The profiles are more oscillatory in Fig. 2 (a) with regard to those in Fig. 2 (b). However, several strong or weak couplings occur temporally in both of the intervals as the big values in the time courses of ($z(\tau)$) indicate in Figs. 3 (a) and (b). Here, a distinguishing property is that the coupled entropy indices do not come close to 1 in the inter-ictal interval. Moreover, no formations similar to these which occur nearly 20 minutes before the imminent seizure onset in Fig. 3 (a) can be found in those profiles for the inter-ictal interval; see, Figs. 2 (b) or 3 (b). Furthermore, similar pre-ictal formations occur before the known seizures of the Pat03 as exemplified in Figs. 4 (a) and (b) where the data are continuous for 2 h in each plot. It should be noted that the temporal couplings; namely, the pin-like topographies covering a few windows are not strong during the seizure indicated in Fig. 4 (b) with respect to those in Fig. 3 (a). However, the values of the indices are close to 1 with broad signals in all. Moreover, various pin-like formations occur also after the seizures for nearly 20 minutes. Therefore, it may be claimed that the spontaneous occurrences of individual strong or weak couplings are not decisive for the initiation of a seizure, where an important criterion is the emergence of a pre-ictal stage as stressed in [20-21].

In summary, a pre-ictal stage originates nearly 20 minutes before the first indicated seizure (Fig. 5 (a), which is the same as Fig. **3** (a) with different time domain) and it evolves intermittently through several coupling epochs, each of which continues nearly one minute covering five or six windows, usually. It is reported in [17 and 18] that, the interdependences can be detected for up to 30 seconds prior to a seizure. Also, see [14-16], and [20] and [21].

The indices come very close to each other around 1 during the seizure terms. The transitions from an ictal state to inter-ictal state also continue for nearly 20 minutes with various strong or weak coupling terms as in the pre-ictal stage and the index values decrease after the offsets; namely, in the post-ictal stage, to their log-run or inter-ictal averages. Thus, a cycle for a seizure is completed. Similar cyclic behavior consisting of pre-ictal, ictal and post-ictal states accompany all of the seizures of the patient (see, Figs. 4 (a) and (b) for example).

It is known that various focal seizures begin in a restricted brain region and either remain localized or spread to the adjacent cortex, as considered in the Introduction section. Let us focus on 3 (a) or 5 (a) to inspect the approach, onset and offset of the first seizure. The pre-ictal stage is presumably originated (due to an unknown biological reason) at about $\tau=1.45$ h where the entire entropy index values decrease for a few minutes. This means that the in-focus or out-focus contact positions are temporarily coupled. The couplings are weakened in the next five or six minutes. Afterwards, the strength and occurrence frequency of the couplings increase intermittently with time and the tops of the pin-like topographies ascend towards 1. All of the contact sites are coupled at about $\tau=1.65$ h, i.e., a few minutes before the onset of the seizure which occurs at about $\tau=1.7$ h; where the maximum values of the indices with $k=1-6$ are nearly the same. During the seizure, integrated neural activity reaches a climax, where all of the indices have nearly the same value which is very close to 1 and EEG signals broaden. Apparently, more brain sites are coupled with increasing strengths while approaching to the seizure onset. It may be claimed that the spread of the seizure is not a continuous process but an intermittent one since strong couplings occur intermittently. The relaxation of the indices is gradual and accompanied by numerous couplings, which are also intermittent. Figs. 5 (b) and (c) show the first and second order successive time differences (approximate time derivatives) of the indices with the same time domain of Fig. 5 (a), where the intermittent couplings are clear. Therefore, not only the indices but also their time differences are coupled at several time epochs. Moreover, the brain electricity may be unique and locally modulated when close to or during a seizure. (Various different suggestions on the subject may be found on page 198 in [39].) This subject will be held while investigating other cases for different patients and further discussed in the last section.

Let us now treat the distributions of the index values for the data from the Pat03 in the ictal interval (Fig. 6 (a)) or in the inter-ictal interval (Fig. 6 (b)), where the aim is to search for statistical evidence about the electrical couplings between different brain sites. The plots for the data from the inter-ictal interval are evidently parabolic in logarithmic scale (or Gaussian in linear scale) whereas those for the ictal interval depict important deviations from Gauss, where the deviations are outstanding especially for the data from the out-focus electrodes, with $k=4-6$. This shows that the out-focus sites are strongly affected by the epileptogenic activities.

Note that the modes of the ED plots of the index values occur at the intermediate values which match the values for the pre-ictal and post-ictal stages. Hence, these stages can be predicted if the modes of the ED plots of the index values show certain deviations from Gauss. Conclusively, the ED plots of the entropy index values can be utilized to distinguish between the inter-ictal data and ictal data or more specifically, the inter-ictal period and the pre-ictal period (see, [20-21]).

The time series of the mutual arithmetical differences of the in-focus ($k=1-3$) or out-focus ($k=4-6$) index values are displayed in Figs. 7 (a) and (b), respectively which cover the first seizure. Note that the differences remain close to each other around zero for a few minutes after the seizure offsets and depart gradually afterwards. This behavior can also be monitored during the other seizures of the Pat03 (Fig. 7 (c)) and the seizures of several other patients as will be considered in further cases.

The over-all behavior of these electrode differences can be understood in terms of the distributions of their values as displayed in Figs. 8 (a) and (b). Inspection of these figures shows that the in-focus and out-focus electrodes are more strongly correlated in the ictal-interval than in the inter-ictal interval since the modes occur at zero in Fig. 8 (a). Moreover, the heights of these modes are nearly the same for all of the electrodes which indicate a strong collective behavior where the heights are much bigger than the number of the windows for the seizures. This is because numerous very strong couplings occur in the pre-ictal and post-ictal stages as well as in the ictal stages. However, the index values come out very close to 1 for these strong couplings only during the seizures, which is a distinguishing property for the onsets. Furthermore, all of the out-focus electrodes and only two of the three in-focus electrodes (k=1 and k=3); namely, FE1 and FD1 (Table I and Fig. 1) are strongly correlated in the inter-ictal interval since the highest and narrowest peak around zero comes out for these electrode differences ($D_{FE1,FD1}$) in Fig. 8 (b). Hence the total duration of the $D_{FE1,FD1}$ is the longest in-between the in-focus or out-focus couplings during the inter-ictal interval. Whereas, all of the out-focus electrodes are strongly coupled in both of the intervals where the couplings are stronger in the ictal interval, Figs. 8 (a) (also, Fig. 7 (c)). Note that, the plots for $D_{1,2}(\tau)$ and $D_{2,3}(\tau)$ are nearly anti-symmetric about zero. This is because, the time profile of $\chi_1(\tau)$ is similar to that of $\chi_3(\tau)$; namely, ($\chi_1(\tau) \leftrightarrow \chi_3(\tau)$) where the symbol (a↔b) is used to designate the similarity (or coupling) between a and b. Two trajectories are said to be similar if their time courses are the same up to the minor fluctuations, in this work. Hence, the ED plot for $D_{1,3}(\tau)$ depict a narrow peak around zero since $D_{1,3}(\tau)=\chi_1(\tau)-\chi_3(\tau)\leftrightarrow 0$ due to Eq. (4). Therefore, ($\chi_1(\tau)-\chi_2(\tau)$) is similar to ($\chi_3(\tau)-\chi_2(\tau)$) and $D_{1,2}(\tau)\leftrightarrow D_{3,2}(\tau)=-D_{2,3}(\tau)$. Moreover, the in-focus electrode k=1 is strongly coupled with all of the out-focus electrodes in the ictal interval whereas the in-focus electrode k=3 is strongly coupled with all of the out-focus electrodes in the inter-ictal interval (not shown). Hence, the strength or durations of the synchronous electrode interdependence may vary with the time.

The distributions of the values of the first or second order temporal differences are (Eqs. (6) or (7), respectively) exponential and symmetric about zero for each electrode in the ictal interval or in the inter-ictal interval (not shown). This subject will be discussed in detail in further sections.

*CASE 2: Pat14 with Fronto/temporal Epilepsy*     Only three of four indicated seizures are detected in this case (Fig. 9 (a)) where the first indicated seizure which occurs a few minutes after an abrupt decrease in the indices (for a narrow signal) is missed (Fig. 9 (b)). During this or other downward pin-like behaviors in the ictal and the inter-ictal data, the electrodes show intermediately strong cross-correlations as displayed in Fig. 9 (c). Moreover, a diversity of EEG signals from narrow to broad signals is generated at different times in this case. (Note that there are no downward pin-like patterns for narrow signals in the previous case.)

The entropy profiles from the in-focus electrodes have big magnitudes and fluctuate with small magnitudes around nearly 0.9 whereas these from the out-focus positions fluctuate with big magnitudes around nearly 0.85 in the ictal interval (Figs. 9 (a) and (b)). The values of the oscillations are above 0.8 also in the inter-ictal interval aside from the frequently occurring downward pin-like behaviors (see Fig. 9 (c), for a short time domain).

Durations of the pre-ictal and post-ictal states are nearly 30 minutes or longer, as can be seen in Fig. 9 (c) where the data are continuous for 3 h and two ictal cycles consisting of pre-ictal, ictal and post-ictal stages are included. The profiles are inter-ictal-like up to τ~4.6 h where weak couplings occur between the electrodes and the entropies little increase. The following pre-ictal state for the onset at τ~4.8 h involves several weak synchronizations and these interdependencies evolve into strong ones during the seizure which starts and terminates abruptly. The post-ictal state lasts for nearly 30 minutes with numerous strong and synchronized interdependencies. The adjacent inter-ictal state also lasts for about 30 minutes, up to τ~6.0 h while strong couplings start which become stronger at τ~6.2 h. Thus, the pre-ictal state for the next seizure which occurs at τ~6.6 h initiates. Note that the values of the entropy indices are bigger in the considered post-ictal stages with respect to these in the inter-ictal or pre-ictal stages.

The first and second order time differences of the individual indices also display the couplings (upward or downward pin-like topographies) in the ictal interval and the inter-ictal interval (not shown). Moreover, the ED plots of the values of these differences are symmetric and exponentially decreasing around zero as in the previous case for the Pat03 (not shown).

On the other hand, the ED of the ictal and inter-ictal index values are characteristically different (as in the previous case) with some noisy behaviors in both of the ED (in this case), as can be seen in Figs. 10 (a) and (b). The noisy behaviors and deformations from Gauss in Fig. 10 (a) are due to the pre-ictal and post-ictal states as discussed in the previous case whereas these in the data from the inter-ictal interval can be attributed to the time lengths of the patient's pre-ictal or post-ictal stages, both of which are nearly 30 minutes or longer. Thus, various data from the pre-ictal or post-ictal epochs might have been mistakenly counted within the one-hour-data-segments of the inter-ictal EEG in [32]. Namely, various pre-ictal states manifest weak or intermediately strong couplings without leading to completed seizure onsets, in the time epochs far from the experienced seizures and these data are counted for the inter-ictal interval. However, the ictal and inter-ictal intervals can be clearly distinguished in terms of the ED plots since these plots, especially for the out-focus electrodes from the ictal interval show important deformations with respect to those from the inter-ictal interval as in the previous case.

Moreover, the maximum values of the indices of this patient are very close to 1 (see, Fig. 10 (a) and (b) also Figs. 6 (a) and (b), 14 (a) and (b), 23 (a) and (b)) whereas the minimum values are 0 due to various narrow signals. Hence very broad and narrow brain signals are generated in this case.

Various simultaneous couplings between the local electrodes of the Pat14 can be utilized to detect seizure cycles as considered in the previous case. Yet, the focus will be on the simultaneous couplings between the non-local electrodes, here; see, Figs. 11 (a) and (b). Although the cyclic behavior is not clear, the ability of the electrode differences to capture the seizure term is obvious in Fig. 11 (a). Moreover, the differences $D_{1,5}(\tau)$ and $D_{2,5}(\tau)$ (the dotted lines in different colors in Fig. 11 (a)) follow each other closely in the given time domain. This feature implies that $D_{1,2}(\tau) \leftrightarrow 0$ due to Eq. (4); namely, the contact positions at k=1 and k=2 are strongly correlated: $\chi_1(\tau) \leftrightarrow \chi_2(\tau)$ thus, $(\chi_1(\tau)-\chi_5(\tau)) \leftrightarrow (\chi_2(\tau)-\chi_5(\tau))$ and $D_{1,5}(\tau) \leftrightarrow D_{2,5}(\tau)$.

The related cyclic behavior can be inspected in Fig. 11 (b) which shows the time courses of a different set of non-local electrode couplings. The first and second integrated behaviors in the profiles towards 0 at τ~3.2 h and τ~3.3 h, respectively indicate the pre-ictal stage. These couplings are very strong during the seizure and weakened afterwards. They continue for several minutes in the post-ictal stages as can be inspected in Fig. 11 (c) for different seizure terms, where all of the time profiles of the non-local couplings are designated by dotted lines.

The ED plots in Fig. 12 (a) and (b) show that several simultaneous non-local cross-correlations are weak but they follow each other closely through the ictal interval possibly with a time phase difference; see, also Fig. 11 (a). (Please note that the frequency or time phase domain analyses are kept beyond the scope of this contribution.) Figs. 12 (c) and (d) are for the cross-correlations of local electrodes which depict that the cross-correlations may vary with time. For example, the $D_{1,3}$ follows the $D_{2,3}$ or $D_{1,2}$ in different time windows and with different magnitudes.

Relying upon the predictions made in the previous paragraph, it may be remarked that the unique brain electricity may be modulated differently at different times in different brain sites, but in coordination with some other site(s). Because, the partnerships in the brain couplings continue for long times yet the partners may be exchanged temporarily.

*CASE 3: Pat20 with Tempo/Parietal Epilepsy*   The data registration from the Pat20 in the ictal interval (the seventh one-hour-data-block; 020322aa_0054_k with k=1-6 in [32]) is interrupted (at i=837632 with k=1-6) for several minutes (Fig. 13 (a) at τ~6.9 h) and no reason for the disconnection is given in [32]. The brain states might have been affected radically if some medical treatment were performed during the disconnection. For the effects of medications such as drugs and anesthetics such as nitrous oxide on the brain states, or several examples of the related Shannon entropy applications, see [53-56] and [36], respectively and the references therein. Whatsoever this situation is, three out of four indicated seizures are detected in this case. Yet, numerous false alerts are produced due to the synchronized behavior in the data except those from the electrode k=4 as exemplified in Fig. 13 (b) where the data are continuous for 3 h. In the same figure, the pre-ictal stage for the first seizure, which is undetected, is very long and consisting of several sub-stages with various durations. At $\tau \sim 0.2$ h, the index values decrease collectively. Afterwards, the values increase and various couplings occur between the contact positions except at k=4. These oscillations continue till $\tau \sim 0.7$ h where a new collective decrement similar to the previous one occurs. This time, the separation between the profiles, from the contact position at k=4 and others, are obvious. The tops of the pin-like topographies increase gradually and approach 1 very closely at $\tau \sim 1.6$ h where a seizure might have been expected to onset as $z(\tau)$ indicates. However, the values decrease abruptly and this state continues for nearly 20 minutes till $\tau \sim 1.9$ h where an onset is realized. Therefore, numerous false attempts for an onset can be found in both of the ictal or inter-ictal data from this patient. (See, Fig. 3 (a) for two similar incomplete formations before the experienced seizure of a different patient.)

It can be observed in Fig. 13 (b) and (c) that the onset and offset of the seizure at $\tau \sim 1.9$ h are abrupt. Moreover, the electrode positions at k=1 and k=6 or k=2-5 are coupled strongly right after the offset for nearly 5 minutes as depicted in Fig. 13 (c). The electrodes change couples afterwards and this epoch continues for nearly 20 minutes. Various false inclinations for a seizure onset might have been observed in the several epochs of the ictal data from this patient (for example, between $\tau \sim 2.6$ h and $\tau \sim 2.9$ h).

The couple exchanges can be monitored in the ED plots of the index values from the ictal interval and the inter-ictal interval (Figs. 14 (a) and (b)) where the collective phenomena are clear for big values of the indices from both of the intervals. Note the strong coupling between the electrodes k=1 and k=6 where this partnership continues throughout the ictal interval and inter-ictal interval.

The time series for the time differences also exhibit the pin-like topographies for the couplings and the ED plots of their values come out symmetric and exponentially decreasing around zero. The local or non-local spatial couplings may run with the same partners in several epochs and the couples may be changed in others. Moreover, these differences may be utilized to detect the same seizure onsets which are captured by means of the marker. Thus, further investigations of the spatiotemporal differences of the electrodes do not provide new information in this case, because they are obtainable using different methods as shown in the previous analyses.

However, it should be noted that the aforementioned (the first paragraph in this CASE) disconnection occurs at $\tau \sim 6.9$ h but the index values decrease to zero much earlier than the disconnection; namely, at $\tau \sim 6.4$ h and this behavior continues till the disconnection. The index profiles with k=1-6 decrease to zero much earlier than the disconnections occur also in the data from several patients in different cases. Hence, there may be a causal relation between the narrowness of the EEG signals before the disconnections and the reason(s) for the disconnections. This subject is further discussed in the last section.

*CASE 4: Pat11 with Parietal Epilepsy*   The entropy indices from the Pat11 are obviously non-stationary as displayed in Fig 15 (a). The indices from the continuous first two one-hour-data-blocks have small values and the seizure term at $\tau \sim 1.05$ h is undetected by the current method. However, both of the two indicated seizures in the continuous data for the following 4 hours (2 h<$\tau \leq$6 h) are captured. Yet, the last seizure which occurs in the last continuous data set (6 h<τ≤8 h) is missed. As a result, two of the four indicated seizures are detected in this case.

It should be noted that the patterns of the profiles for the second and third seizure terms are similar to each other and thus there are three sets of continuous data and three types of seizure profiles in this case (Figs. 15 (a)-(c)). The first set of the continuous data deliver nearly homogenous index profiles where no pre-ictal, ictal or post-ictal stages are distinguished. Obviously, the first seizure does not show a cyclic behavior. Whereas, the cyclic interdependence is clear in the second and third seizure terms (Fig. 15 (b)).

The index values gradually increase from τ~2.0 h up to τ~2.5 h and abruptly decrease to the previous levels a little later (at τ~3.0 h). Afterwards, they increase again gradually up to τ~3.1 h and then relax. Both of these eras are clearly not inter-ictal-like but they resemble the pre-ictal states for the imminent two seizures. Thus, they can be taken as failed alerts for a seizure. A clear pre-ictal state for the onset at τ~3.6 h stars at τ~3.3 h where the out-focus index values rise more rapidly than the in-focus ones. This behavior can be inspected also in the pre-ictal stage of the onset at τ~4.8 h. The onsets and offsets of both of these seizures are abrupt and the couplings are not strong in-between them. The intermediate (inter-ictal) stages are depicted at τ~4.0 h and τ~6.5 h both of which continue for nearly half an hour.

The ED plots of the ictal or inter-ictal indices are depicted in Figs. 16 (a) and (b), respectively where the integration is clear for big values. However, the maximum values come out bigger in the ictal interval (for the onsets). Moreover, the plots for the in-focus electrode at k=1 and out-focus one at k=6 are similar in both of the intervals, which shows that the k=1 and k=6 sites are less affected by the other sites. Whereas the same plots for the data from the k=4 position show big variation from one interval to the other and hence it can be claimed that the brain site at k=4 is more effected by the others at the times close to the seizures.

The analyses of the ED plots for the spatial differences of the electrodes show that the in-focus electrodes are strongly correlated in both of the ictal interval and inter-ictal interval, Figs. 17 (a) and (b). Three further strong local or mixed cross-correlations come out in the inter-ictal interval; between the positions at k=4 and k=6, or k=1 and k=6, and k=2 and k=6, respectively. But, the non-local coupling between the sites at k=4 and k=6 weakens in the ictal interval. Please note that numerous time windows involving disconnections or very narrow signals (with zero index values) in the inter-ictal interval are omitted in Fig. 17 (b).

*CASE 5: Pat15 with Temporal Epilepsy*     The first four one-hour ictal data blocks with k=1-6 are continuous which cover two seizures as depicted in Figs. 18 (a) and (b). The pre-ictal state for the first onset starts at τ~0.2 h and continues till τ~0.9 h where the onset is experienced. The post-ictal state is not clear for this seizure. As a result, the pre-ictal state for the second seizure can be hardly detected. The second onset occurs at τ~3.1 h and the post-ictal stage of this seizure continues till τ~3.7 h without a doubt.

Also the next three one-hour-data blocks with k=1,6 are continuous and they cover one seizure (Figs. 18 (a) and (c)) where the indices oscillate with big magnitudes. The oscillations are minor in the time interval between τ~4.8 h and τ~5.2 h which may be taken as the intermediate state. The pre-ictal state starts at τ~5.2 h as distinguished in terms of collective behavior. In other words, the pin-like synchronous interdependencies start at that time and gain importance with time; namely, the maximum values ascend towards 1 and the frequency of the occurrences increase with time. At τ~5.2 h, the integrated behavior reaches a climax, where all of the indices have nearly the same value which is very close to 1. Moreover, the indices drop two times; one during the seizure and one right after the seizure offset. The index values increase abruptly or gradually afterwards till τ~5.85 h and this instant can be taken as the termination of the post-ictal stage which indicates the completion of a seizure cycle. Note that, the electrode couplings continue till $\tau \sim 6.05$ h with big index values which become close to 1 at $\tau \sim 6.3$ h. The time epoch between $\tau \sim 5.85$ h and $\tau \sim 6.05$ h should be considered not as a part of the last post-ictal term but a new pre-ictal state for an unrealized onset at $\tau \sim 6.3$ h. Two more similar inclinations which failed before an onset can be monitored in Fig. 18 (c). The related clinical evidence is not reported in [32].

The last seizure onset occurs at $\tau \sim 8.6$ h in the interval 7.0 h$<\tau \leq$10.0 h which contains continuous data. The index values collectively increase towards 1 with various strength of the couplings several times before the onset; such as at $\tau \sim 7.3$ h and $\tau \sim 7.9$ h. Yet, the strong couplings start at $\tau \sim 8.3$ h which can be taken as the beginning of the pre-ictal stage of the seizure and thus, this stage continues for nearly 20 minutes. An abrupt decrease in the index values occur within the seizure term and these decrements are followed by gradual increments. Hence, the seizure offsets before the completion of the gradual increment term. The values come close to each other around 0.95, that is, a few minutes after the offset and then the post-ictal state occurs and continues till $\tau \sim 7.3$ h.

All of the indicated seizures (four out of four) are detected in this case besides numerous pre-ictal stages of the assumed uncompleted seizure cycles. The pre-ictal or post-ictal transition terms of the realized seizures may last half an hour or longer. Thus, also the ED plots from the inter-ictal interval come out with noise and various deformations from Gauss in the range for the intermediate index values, as those from the ictal-interval in the declared range of the index values; see Figs. 19 (a) and (b). Moreover, the tails of the plots for the big values in Fig. 19 (b) indicate that, the inter-ictal data involve several pre-ictal stages for uncompleted seizure cycles which are manifest also in the time profiles (not shown). Furthermore, the couplings between all of the electrodes, with the exception of the one at k=4 are strong during the ictal-interval with long durations.

The individual behavior of the electrode at k=4 is clear in the ED plots for the spatial differences from ictal interval and inter-ictal interval in Figs. 20 (a) and (b), respectively. Note that no strong cross-correlations involve data from the site at k=4 in both of the intervals. Moreover, the groups of the ED plots for $D_{4,5}$ and $D_{4,6}$ (local) or $D_{1,4}$, $D_{2,4}$ and $D_{3,4}$ (non-local) are similar to each other and the groups of the plots are nearly anti-symmetric about zero in the ictal interval (Fig. 20 (a)). This anti-symmetry is exemplified in Fig. 21 (a) within the three-hour-continuous data which cover the third seizure term as in Fig. 18 (c). This anti-symmetry occurs since the plots of $D_{1,4}$, $D_{2,4}$, $D_{3,4}$, and -$D_{4,5}$ (=$D_{5,4}$) and -$D_{4,6}$ (=$D_{6,4}$) come out similar. In other words, the index from the electrode at k=4 follow a different trajectory than the other electrodes where the differences in the trajectories of the indices from the electrodes with k≠4 can be neglected (the central group in Fig. 20 (a)). As a result, the plots of $D_{k,4}$ with k≠4 come out similar. Therefore, all of the brain sites except at k=4 couple strongly in the ictal interval of this patient.

Fig. 21 (b) is the same as Fig. 21 (a) for a shorter time range and for all of the electrode couplings, where the following properties are inspected: All of the electrodes are coupled strongly for a while which implies that the brain electricity is unified at every site in these time terms. Moreover, the seizure offsets whilst the electricity is unique. The couplings become strong once again at $\tau \sim 6.8$ h with a longer duration. Fig. 21 (c) covers the last seizure where a similar behavior to the one described in the last lines of the previous paragraph is manifest yet the offset occurs during the strong couplings of the brain sites.

The time profiles of the successive time differences of the individual recordings display the temporal couplings and their values are distributed symmetrically and exponentially around zero as in the previous cases (not shown).

*CASE 6: Pat09 with Temporo/Occipital Epilepsy* The continuously updated index values of the patient oscillate with big magnitudes in ictal interval (Fig. 22 (a)) where, four out of five of the indicated seizures are detected and several false alerts are obtained. The beginning of the pre-ictal stage for the first seizure ($\tau \sim 1.05$ h) is not clear. Weak couplings with short durations are inspected at the beginning of the first one-hour-data-blocks with k=1-6. These weakly synchronized interdependencies become stronger and the event frequencies increase with the time. The post-ictal stage continues nearly 20 minutes after the offset and several pin-like formations appear afterwards till $\tau \sim 2.0$ h, which is the end of the first set of continuous one-hour-data-blocks. The data for 2 h<$\tau \leq$6 h are also continuous and they involve two seizures (Fig. 22 (a)) which are accompanied by clear pre-ictal states with the time durations of nearly 30 minutes. These pre-ictal states can be distinguished from the intermediate-states in terms of the increasing number of strong couplings with time. Whereas, the time durations of the post-ictal states of these seizures are shorter than these of the pre-ictal ones.

The most interesting feature in the results from the patient is the strong couplings between all of the electrodes at 8.1 h<$\tau$<8.4 h which continue for nearly 20 minutes (Figs. 22 (a) and (b)). Nearly at the beginning of the eighth one-hour-data-blocks (010906ea_0070_k with k=1-6), the entropies of the individual recordings follow each other very closely for about nearly 20 minutes yet no seizure is indicated in [32] for the current interval. Note that a similar formation occurs also in the data from the inter-ictal interval as displayed in Fig. 22 (c), where the data are continuous along the interval.

The ED plots of the index values from the ictal interval (Fig. 23 (a)) and inter-ictal interval (Fig. 23 (b)) are capable of distinguishing these intervals, clearly. Note that the electrodes do not exchange partners in the regime for small index values and the out focus electrode at k=6 is coupled with the in-focus electrodes in both of the intervals. The maximum values of the indices are very close to 1 with very broad or homogeneous signals as in CASE 2.

The cross-correlations between the electrodes at k=5 and k=6 are strong in the ictal and inter-ictal interval where the ED plots for these correlations and that for the group of $D_{1,5}$, $D_{2,4}$, $D_{3,4}$ and $D_{3,5}$ are anti-symmetric. The temporal differences of the individual recordings do not provide additional information since they are similar to those from the other patients. These two subjects are further discussed in the next section.

In summary, nearly 700 million ictal or inter-ictal EEG data intracerebrally recorded from 20 patients with different ages, genders, origins of the epilepsy, seizure types and electrode types or insertion places are investigated in terms of the continuously updated entropy indices $\chi_k(\tau)$ with k=1-6. A common property predicted in the results, which are exemplified in detail for 6 patients, is that the index values increase and then decrease several times before the imminent seizures and reach at locally maximum values close to 1 during the onsets. The characteristic features of these similarities are: 1) The indices from the ictal interval may fluctuate about different values for different k with weak couplings. This era may be called the intermediate (inter-ictal) state during which no seizure is expected to occur soon. 2) When the number of the couplings per a given time interval and their strength increase with the time, this means that the patient is approaching an onset of a seizure. This era stars 20-30 minutes before the onset and it may be called the pre-ictal state. If the tops; namely, the local maximum values of the entropy indices increase and the synchronizations cover more univariates from different brain sites then the onset is expected to occur within a few minutes. 3) The detected seizure onsets are accompanied by synchronized and strong couplings in the entropy indices with values very close to 1. This means that the brain signals are unified at several sites and broad during these seizures. 4) The indices relax after the offsets gradually in most of the cases or abruptly in forms downward pin-like formations in a few cases; as in Case 5, for example). This era may be taken as the post-ictal stage which continues for 20-30 minutes usually.

It may be remarked that the synchronization in the EEG signals can be monitored as a parallelization in the time profiles of the entropy indices or their successive time differences, $\chi'_k(\tau)$ or $\chi''_k(\tau)$. The time courses of variation of these parallel behaviors are investigated also in terms of the electrode couplings in order to better understand the underlying mechanisms.

**Discussion and conclusion**: The presented approach provides a reasonable description of various spatiotemporal similarities in between the brain states before, during and after the captured seizures. These similarities are predicted to be cyclic spatiotemporal synchronization (coupling) of the brain sites where the spatial couplings may be local; namely, between two (or three) in-focus electrodes or between two (or three) out-focus ones, or non-local; namely, between numerous in-focus and out-focus electrodes. Moreover, the durations of these partnerships or the partner sites may change with time.

The patterns of the integrated neuronal behavior observed in the entropy indices, as well as in the spatiotemporal differences (approximate derivatives) of the indices, come out patient-specific, which may be related to psychological fluctuations such as, vigilance (wakefulness, drowsiness or sleep), attention and anxiety in a patient. These interrelationships are found to be weak and infrequent in the inter-ictal intervals and when they intermittently evolve into strong and frequent ones then this stage is distinguished as a pre-ictal state. Thus, the epileptogenic processes are predicted to be intermittent. Similar results are reported in [58] where long-term intracranial recordings of five patients by employing a measure of phase synchronization are investigated.

Furthermore, it is observed that extremely strong couplings occur between the brain sites during the predicted seizure onsets and the transitions from these onsets to post-ictal or inter-ictal stages also follow patient-specific patterns in the time profiles of the entropy indices and their spatiotemporal differences. As a result, all of the patterns in these time profiles are clearly patient dependent. Therefore, not only the epileptogenic processes but also the relaxation processes are patient-specific, intermittent and cyclic where a cycle for the entire ictal process consists of three stages: (pre-ictal) preparation for, (ictal) realization of and (post-ictal) recovery from a seizure.

However, the distributions of the values of the first order successive time derivatives of the indices (Eq. (6)) seem patient independent as displayed in Fig. 24 where all of the available data in [32] from 20 patients in ictal interval or inter-ictal interval are used. (The total number of the utilized time windows with no disconnection or non-zero entropy values in Fig. 24 is 234,879.) It should be noted that the plots of these distributions are symmetric and exponentially decreasing for each of the patients in ictal interval or inter-ictal interval as considered in the previous section for various cases. Moreover, the ED plots of the values of the second order successive time derivatives (Eq. (7)) come out similar to those in Fig. 24. It can be claimed that the distribution curves of $\chi'_k(\tau)$ and $\chi''_k(\tau)$ with k=1-6 are similar not only for focal epilepsies but for all of the epileptic and possibly healthy brains. (The tails, for the big magnitudes, will come out presumably clear for healthy brains).

Furthermore, the presented approach can be considered successful since nineteen out of the twenty six studied seizures are captured as discussed in detail in the previous section. This constitutes a more than 70 % success rate. A similar success rate is achieved also in [58].

The types or insertion places of the used in-focus or out-focus electrodes are different (Fig. 1). Hence, the capability of the marker $z(\tau)$ (Eq. (4)) to capture the seizure onsets may change from one case to the other. Moreover, a smoothed version of $z(\tau)$ (namely, the application of a known smoothing function on $z(\tau)$) or averages of $z(\tau)$ over a fixed number of the backward or forward time windows at a time τ, may yield more precise detections where some big powers of $|\chi_k(\tau)|$ or $|\chi_{k'}(\tau)|$ can also be utilized in Eq. (4).

It can be expected that the accuracy of the results may depend on the length of the time windows used due to the non-stationary character of the epileptic EEG signals.

*Length dependence*:   Various analyses of magnitude square coherence and phase coherence, entropy or entropy and complexity show that the results depend on the data length; see Table 1 in [59] and the results in [60] or [61], respectively.

In the results of this work presented above, the window length is 10 seconds. Thus each window involves 2560 data per electrode (k), since the sampling rate is 256 Hz. Therefore, numerous bins in the ED of the EEG voltages ($Q_k(i)$ in Eq. (1)) come out empty ($Q_k(i)=0$) since the bin size is 1 μV and the range of the voltages is from -10,000 μV to 10,000 μV or broader. Hence, the window length may be crucial for the results, especially when a spiky behavior or narrow or broad signal is dominant in a given epoch. In this case, longer time windows can be used in order to reduce the effect of these irregularities on the results and to avoid abnormal values for the indices.

The values of the entropy indices decrease to zero before various indicated disconnections occur during the data registration in [32]. For example, in CASE 3 in the Results section, the index values of the Pat20 in the ictal interval descend towards zero at τ~6.4 h (Fig. 13 (a)) and the values remain zero till a disconnection occurs in the data registration nearly 30 minutes later; namely, at τ~6.9 h (i=837632 with k=1-6). This appearance is observed in the EEG from the same patient in the inter-ictal interval or from different patients in both of the intervals. Hence, there may be a causal relationship between several irregular behaviors in the EEG signals and the simultaneous clinical symptoms of the relevant patients before the disconnections. These symptoms are not reported in [32]. Moreover, the effect of these irregular data may be reduced in the entropy estimations when longer time windows are used as exemplified in Fig. 25(b).

Figs. 25 (a)-(c) are obtained using time windows of length 100 seconds in the presented approach. In Figs. 25 (a) and (b), the index values come out smaller than those in Fig. 13 (a)-(c) by a factor of nearly 1.3 (~10.15/7.85=~ln(25600)/ln(2560), see Eq. (3)). The time domain of Fig. 25 (a) is the same as that of Fig. 13 (b) where the following three features can be clearly observed: the ictal cycle of the experienced onset at τ~1.9 h, several unsuccessful inclinations for a seizure before the onset and the individual behavior of $\chi_4(\tau)$ (thick line in dark cyan in Fig. 25 (a)). The individual behavior of $\chi_4(\tau)$ can also be observed in Fig. 25 (b) where the data contains a disconnection as discussed in the previous paragraph and Results section (CASE 3: Pat20). It must be noted that the index values are non-zero also after τ~6.4 h till the occurrence of the disconnection at τ~6.9 h in Fig. 25 (b).

The successive differences of the indices ($\chi_k'(\tau)$) with the newly used longer time windows (Δτ=100 seconds in Eq. (6)) are displayed in Fig. 25 (c) where the time domain includes the first indicated seizure of the Pat20, at τ~1.8 h which was undetected previously.

The patterns obviously depend on the length of the time windows; however, the characteristic features of the results (for example, the cyclic topographies of the ictal patterns) remain invariant within reasonable limits for the length of the time windows.

*Couplings and brain electricity*:   If the entropy indices from the brain sites at k and k' are similar ($\chi_k(\tau) \leftrightarrow \chi_{k'}(\tau)$) in a time interval then these sites are called coupled or synchronously interdependent in this work as referred to above. Thus, the four fundamental states of an epileptic brain; namely, the inter-ictal, pre-ictal, ictal and post-ictal states can be distinguished in terms of the simultaneous interrelationships. The couplings are found to be infrequent and weak in an inter-ictal stage, as in [19] for healthy brains. The event frequency and strength of the couplings usually increase with time in a pre-ictal state [20 and 21] and they decline after the offsets for a post-ictal stage. In the meantime the strengths reach a climax for a few minutes for a seizure. If the number of the coupled brain sites also increase in a pre-ictal state then the brain electricity and possibly the seizure may be taken as spreading.

Various similarities between the brain electricity recorded from different sites are detected in this contribution for different cases and thus the spread of the brain electricity is discussed. However, these results could not be associated with the seizure types because no further information about the individual seizure types is available in [32]. In other words, the predictive power of the presented approach depends on the number and strength of the brain couplings at a time. Thus, if a seizure onset occurs either with weak or a few brain couplings then it could not be detected. It is not known which of the noted seizure types [32]; namely, simple partial (SP), complex partial (CP) or generalized tonic-clonic (GTC) onsets fits which of the investigated seizures. The presented approach might provide statistical evidence between a seizure type and brain couplings if individual seizure types were known.

Let us consider the aforementioned (two paragraphs above) similarity between the two entropy index trajectories, $\chi_k(\tau)$ and $\chi_{k'}(\tau)$. It is clear that, if $(\chi_k(\tau) \leftrightarrow \chi_{k'}(\tau))$ then $(\chi_k(\tau)-\chi_{k''}(\tau)) \leftrightarrow (\chi_{k'}(\tau)-\chi_{k''}(\tau))$. Thus, $D_{k,k'}(\tau) \leftrightarrow 0$ and $D_{k,k''}(\tau) \leftrightarrow D_{k',k''}(\tau)$. Such a situation was considered in CASE 1 for the Pat03 in the Results section where $D_{1,3}(\tau) \leftrightarrow 0$ and $D_{1,2}(\tau) \leftrightarrow -D_{2,3}(\tau)=D_{3,2}(\tau)$ for 1 h<$\tau$< 2 h which covers a seizure with unknown type, (Fig. 7 (c)). Another example is given in CASE 2 for the Pat14 as depicted in Fig. 11 (a) where $D_{1,5}(\tau) \leftrightarrow D_{2,5}(\tau)$ (dotted lines) as a result $D_{1,2}(\tau) \leftrightarrow 0$ as shown in Fig 12 (a).

Thus, if two brain sites k and k' are coupled strongly then $D_{k,k''}(\tau)$ and $D_{k',k''}(\tau)$ with k''≠k'≠k, follow each other very closely. The stronger is the coupling between the brain sites at k and k', the smaller is the difference between the trajectories of $D_{k,k''}(\tau)$ and $D_{k',k''}(\tau)$ in a given time epoch. In this case, the ED plots for $D_{k,k''}(\tau)$, $D_{k',k''}(\tau)$ and $D_{k,k'}(\tau)$ coincide with each other about the zero. Therefore, the modes of the ED plots of $D_{k,k''}(\tau)$ and $D_{k',k''}(\tau)$ are far from or close to each other if that of $D_{k,k'}(\tau)$ is far from or close to zero, respectively. See, the trajectories of various spatial differences in Figs. 7 (a) and (b), 11 (a)-(c), and 21 (a)-(c) and their ED plots in Figs. 8 (a) and (b), 12 (a)-(d), and 20 (a), respectively.

# APPENDIX

Suppose that the number of the samples X(i) in a finite set is N. Let us consider Eq. (1) in the main text with P(X(i))=Q(X(i))/N where P(X(i))=P(i) and Q(X(i))=Q(i) are the normalized and un normalized empirical distributions of X(i), respectively and $1 \leq i \leq N$;

$$U = -N\sum_{i=1}^{N} P(i)\log(NP(i)) \tag{A1}$$

or $$U = -N\sum_{i=1}^{N}\big(P(i)\log(P(i) + P(i)\log(N)\big) \tag{A2}$$

or $$U = NS - N\log(N) \tag{A3}$$

since P is normalized, $\sum_{i=1}^{N} P(i)=1$. Thus,

$$U/N = S - \log(N) \tag{A4}$$

or $$U/(N\log(N)) = S/\log(N) - 1 \tag{A5}$$

or $$1 + U/(N\log(N)) = S/\log(N) = \chi \tag{A6}$$

where $\chi$ is the same as the index in Eq. (2) in the main text.

ACKNOWLEDGEMENT

The author is thankful to the University of Freiburg for their kindness in giving permission to investigate their database and use all of the related material.

**REFERENCE LIST**

[1] F. H. Lopes da Silva, EEG analysis: theory and practice; in: E. Niedermeyer, F. H. Lopes da Silva, eds. Electroencephalography: basic principles, clinical applications and related fields, Baltimore: Williams and Wilkins, 3rd ed. 1993.
[2] F.H. Lopes da Silva, S. van Leeuwen, A. Réemond, Handbook of Electroencephalography and Clinical Neurophysiology, Vol. II: Clinical Application of Computer Analysis of EEG and Other Neurophysiological Signals, Elsevier, Amsterdam, 1986.
[3] F. H. Lopes Silva, A. Hoeks, T. H. M. T. van Lierop, C. F. Schrijer and W. S. van Leeuwen, Confidence intervals of spectra and coherence functions—Their relevance for quantifying thalamo-cortical relationships; in Quantitative Analysis of the EEG M. Matejcek and G. K. Schenk, eds. AEG Telefunken, Constanz, 1975.
[4] P. Kellawou and I. Peterson, Automation of Clinical EEG, eds. New York: Raven, 1973.
[5] S. Faula, G. Boylanb, S. Connollyc, L. Marnanea, G. Lightbodya, An evaluation of automated neonatal seizure detection methods, Clinical Neurophysiology **116**, 1533, 2005.
[6] J. D. Bronzino, M. L. Kelly, C. T. Cordova, N. H. Oley, and P. J. M., Utilization of Amplitude Histograms to Quantify the EEG Effects of Systemic Administration of Morphine in the Chronically Implanted Rat, IEEE BME-**28**, 673, 1981.
[7] J. S. Barlow, Computerized Clinical Electroencephalography in Perspective, IEEE BME-**26**, 377, 1979.
[8] A. S. Gevins, C. L. Yeager, S. L. Diamond, J.-P. Spire, G. M. Zeitlin, and A. H. Gevins, Automated analysis of the electrical activity of the human brain (EEG): A progress report, Proc. IEEE, vol. 63, 1382, 1975.
[9] Y.U. Khan and J. Gotman, Wavelet based automatic seizure detection in intracerebral Electroencephalogram, Clinical Neurophysiology **114**, 898, 2003.
[10] C. D. Binnie, B. G. Batchelor, P. A. Bowring, C. E. Darby, L. Herbert, D. S. L. Lloyd, D. M. Smith, G. F. Smith, and M. Smith, Computer-assisted interpretation of clinical EEGs, Electroencephalogr. Clinical Neurophysiology, **44**, 575, 1978.
[11] H. W. Shipton, EEG analysis: A history and a prospective, Annu. Rev. Biophys. Bioeng., **4**, 1, 1975.
[12] N. V. Thakor and S. Tong, Advances in quantitative electroencephalogram analysis methods. Annu. Rev. Biomed. Eng. **6**, 453, 2004.
[13] I. Osorio, M-G. Frei, S. B. Wilkinson. Real-time automated detection and quantitative analysis of seizures and short-term prediction of clinical onset, Epilepsia, **39**, 615, 1998.
[14] T. Kreuz, F. Mormann, R. G. Andrzejak, A. Kraskov, K. Lehnertz, P. Grassberger, Measuring synchronization in coupled model systems: A comparison of different approaches, Physica D **225**, 29, 2007.
[15] F. H. Lopes da Silva, W. Blanes, S. Kalitzin, J. Parra, P. Suffczynski, D. Velis, Dynamical diseases of brain systems: different routes of epileptic seizures. IEEE BME-**50**, 540, 2003.
[16] R. G. Andrzejak, D. Chicharro, C. E. Elger, F. Mormann, Seizure prediction: Any better than chance?, Clinical Neurophysiology **120**, 1465, 2009.
[17] M. Le Van Quyen, J. Martinerie, C. Adam, F. Varela, Nonlinear analyses of interictal EEG map the interdependencies in human focal epilepsy. Physica D, **127**, 250, 1999.
[18] J. Arnhold, P. Grassberger, K. Lehnertz, C. Elger, A robust method for detecting interdependencies: application to intracranially recorded EEG. Physica D, **134**, 419, 1999.
[19] C. Stam, J. Pijn, P. Suffczynski, F. H. Lopes da Silva, Dynamics of the alpha rhythm: evidence for non-linearity? Clinical Neurophysiology **110**, 1801, 1999.

[20] F. Mormann, T. Kreuz, C. Rieke, R. G. Andrzejak, A. Kraskov, P. David, C. E. Elger, Klaus Lehnertz, On the predictability of epileptic seizures, Clinical Neurophysiology **116**, 569, 2005.
[21] F. Mormann, R. G. Andrzejak, C. E. Elger and K. Lehnertz, Seizure prediction: the long and winding road, Brain, **130**, 314, 2007.
[22] M. E. Weinand, L. P. Carter, W. F. El-Saadany, P. J. Sioutos, D. M. Labiner, K. J. Oommen, Cerebral blood flow and temporal lobe epileptogenicity, J. Neurosurg, **86**, 226, 1997.
[23] C. Baumgartner, W. Serles, F. Leutmezer, E. Pataraia, S. Aull, T. Czech, U. Pietrzyk, A. Relic, I. Podreka, Preictal SPECT in temporal lobe epilepsy: regional cerebral blood flow is increased prior to electroencephalography-seizure onset, J. Nucl. Med. **39**, 978, 1998.
[24] P. D. Adelson, E. Nemoto, M. Scheuer, M. Painter, J. Morgan, H. Yonas, Noninvasive continuous monitoring of cerebral oxygenation periictally using nearinfrared spectroscopy: a preliminary report. Epilepsia **40**, 1484, 1999.
[25] P. Federico P, D. F. Abbott, R. S. Briellmann, A. S. Harvey, G. D. Jackson, Functional MRI of the pre-ictal state, Brain **128**, 1811, 2005.
[26] R. Delamont, P. Julu, G. Jamal, Changes in a measure of cardiac vagal activity before and after epileptic seizures, Epilepsy Res., **35**, 87, 1999.
[27] V. Novak, A. L. Reeves, P. Novak, P. A. Low, F. W. Sharbrough, Time-frequency mapping of R-R interval during complex partial seizures of temporal lobe origin. J. Auton Nerv. Syst. **77**, 195, 1999.
[28] D. H. Kerem and A. B. Geva, Forecasting epilepsy from the heart rate signal, Med. Biol. Eng. Comput. **43**, 230, 2005.
[29] E. Pereda, R. Q. Quiroga, J. Bhattacharya, Nonlinear multivariate analysis of neurophysiological signals, Progress in Neurobiology, **77**, 1, 2005.
[30] M.J. van der Heydena, D.N. Velis, B.P.T. Hoekstra, J.P.M. Pijn, W. van Emde Boas, C. W. M. van Veelen, P.C. van Rijen, F.H. Lopes da Silva, J. De Goede, Non-linear analysis of intracranial human EEG in temporal lobe epilepsy, Clinical Neurophysiology **110**, 1726, 1999.
[31] T. Schreiber, Interdisciplinary application of nonlinear time series methods, Physics Reports **308**, 1, 1999.
[32] Freiburg University, http://epilepsy.uni-freiburg.de/prediction-contest/data-download
[33] B. Bein, Entropy, Best Practice & Research Clinical Anaesthesiology **20**, 101, 2006.
[34] P. Viola, N. N. Schraudolph, T. J. Sejnowski, Empirical Entropy Manipulation for Real-World Problems, in Advances in Neural Information Processing Systems 8 (NIPS*96), D. Touretzky, M. Mozer and M. Wasselmo (eds.), MIT Press, 1996; pp.851-857.
[35] A. Isaksson, A. Wennberg, L. H. Zetterberg, Computer Analysis of EEG Signals with Parametric Models, PROCEEDINGS OF THE IEEE, **69**, 451, 1981.
[36] A. Bezerianos, S. Tong S, N. Thakor, Time-dependent entropy estimation of EEG rhythm changes following brain ischemia, Ann. Biomed. Eng, **31**, 221, 2003.
[37] C. E. Shannon, W. Weaver, The Mathematical Theory of Communication, University of Illinois Press, Urbana, 1948.
[38] C. E. Shannon, Bell Systems Technol. J. **27**, 379, 1948.
[39] S. M. Zoldi, A. Krystal, H. S. Greenside, Stationarity and Redundancy of Multichannel EEG Data Recorded During Generalized Tonic-Clonic Seizures, Brain Topography, **12**, 187, 2000.
[40] M. Breakspear and J.R. Terry, Detection and description of non-linear interdependence in normal multichannel human EEG data, Clinical Neurophysiology **113**, 735, 2002.
[41] J. S. Richman and J. R. Moorman, Physiological time-series analysis using approximate entropy and sample entropy, Am. J. Physiol. Heart. Circ. Physiol. **278**, 2039, 2000.

[42] O. Vasicek, A Test for Normality Based on Sample Entropy, Journal of the Royal Statistical Society B, **38**, 54, 1976.
[43] E. J. Dudewicz and E. C. van der Meulen, Entropy-Based Tests of Uniformity, Journal of the American Statistical Association, **76**, 967, 1981.
[44] S. Park, A goodness-of-fit test for normality based on the sample entropy of order statistics, Statistics & Probability Letters, 44, 359, 1999.
[45] R. A.H. Lorentz, On the entropy of a function, Journal of Approximation Theory, **158**, 145, 2009.
[46] J. Timmer, from Freiburg University (private communication).
[47] R. Aschenbrenner-Scheibe, T. Maiwald T, M. Winterhalder, H. U. Voss, J. Timmer, A. Schulze-Bonhage, How well can epileptic seizures be predicted? An evaluation of a nonlinear method, Brain, **126**, 2616, 2003.
[48] M. Winterhalder, B. Schelter, T. Maiwald, A. Brandt, A. Schad, A. Schulze-Bonhage, J. Timmer, Spatio-temporal patient-individual assessment of synchronization changes for epileptic seizure prediction. Clinical Neurophysiology **117**, 2399, 2006.
[49] M. Winterhalder, T. Maiwald T, H. U. Voss, R. Aschenbrenner-Scheibe, J. Timmer, A. Schulze-Bonhage, The seizure prediction characteristic: a general framework to assess and compare seizure prediction methods, Epilepsy Behav., **4**(3), 318, 2003.
[50] T. Maiwald, M. Winterhalder, R. Aschenbrenner-Scheibe R, H. U. Voss, A. Schulze-Bonhage, J. Timmer, Comparison of three nonlinear seizure prediction methods by means of the seizure prediction characteristic, Physica D, **194**, 2004.
[51] B. Schelter, M. Winterhalder, T. Maiwald, A. Brandt, A. Schad, J. Timmer, A. Schulze-Bonhage.: Do false predictions of seizures depend on the state of vigilance? A report from two seizure prediction methods and proposed remedies, Epilepsia, **47**, 2058, 2006.
[52] B. Schelter, M. Winterhalder, T. Maiwald, A. Brandt, A. Schad, A. Schulze-Bonhage, and J. Timmer Testing statistical significance of multivariate time series analysis techniques for epileptic seizure prediction, Chaos **16**, 013108, 2006.
[53] J. Bruhn, L. E. Lehmann, H. Ropcke, T. W. Bouillon, A. Hoeft, Shannon entropy applied to the measurement of the electroencephalographic effects of desflurane, Anesthesiology, **95**, 30, 2001.
[54] I. J. Rampil, A primer for EEG signal processing in anesthesia, Anesthesiology, **89**, 980, 1998.
[55] W. D. Smith, R. C. Dutton, N. T. Smith, Measuring the performance of anesthetic depth indicators, Anesthesiology **84**, 38, 1996.
[56] D. T.J. Liley, K. Leslie, N. C. Sinclair, M. Feckie, Dissociating the effects of nitrous oxide on brain electrical activity using fixed order time series modeling, Computers in Biology and Medicine, *38*, 1121, 2008.
[58] M. Le Van Quyen, J. Soss, V. Navarro, R. Robertson, M. Chavez, M. Baulac, J. Martinerie, Preictal state identification by synchronization changes in long-term intracranial EEG recordings, Clinical Neurophysiology **116**, 559, 2005.
[59] S. Whiting, T. Ning, and J. D. Bronzino, Data length effects on the coherence estimate of EEG, Bioengineering Conference, 1989., Proceedings of the 1989 Fifteenth Annual Northeast, page 93, 1989.
[60] T. D. de Wit, When do finite sample effects significantly affect entropy estimates? Eur. Phys. J. **B 11**, 513, 1999.
[61] M. A. Jiménez-Montano, W. Ebeling, T. Pohl, Paul E. Rapp, Entropy and complexity of finite sequences as fluctuating quantities, BioSystems **64**, 23, 2002.

**TABLES**

|        |       |             | *Frontal* |
|--------|-------|-------------|-----------|
| Pat01  | 15,f  | SP,CP       | G_A4, IH4, IH3; G_D2, IHA1, IH1 |
| Pat03  | 14,m  | SP,CP       | FE1, FE2, FD1; G_F8, G_G8, G_H8 |
| Pat05  | 16,f  | SP,CP,GTC   | G_A3, G_A4, IHB4; G_D1, IHB1, LL1 |
| Pat08  | 32,f  | SP,CP       | FRA2, FRA3, FRC1; G_A8, FLA4, FLC6 |
| Pat18  | 25,f  | SP,CP       | G_A7, G_C5, G_E5; IHC4, IHB4, G_F3 |
| Pat19  | 28,f  | SP,CP,GTC   | G_B5, G_D8, G_G7; G_C1, G_A1, G_H4 |

|        |       |             | *Fronto/Temporal* |
|--------|-------|-------------|-------------------|
| Pat14  | 41,f  | CP,GTC      | FRA1, FRB1, TRB2; TRA1, TRC1, TRC6 |

|        |       |             | *Tempo/Parietal* |
|--------|-------|-------------|------------------|
| Pat20  | 33,m  | SP,CP,GTC   | G_D3, G_B4, G_A2; H_5, FLB4, TLC1 |

|        |       |             | *Parietal* |
|--------|-------|-------------|------------|
| Pat11  | 10,f  | SP,CP,GTC   | G_G3, G_G4, G_E3; G_C8, G_D8, G_F7 |

|        |       |             | *Temporal* |
|--------|-------|-------------|------------|
| Pat04  | 26,f  | SP,CP,GTC   | HR_2, HR_5, TBB1; HR_9, G_D6, G_A1 |
| Pat07  | 42,f  | SP,CP,GTC   | TLA1, TLB2, TLC2; TLA5, TLB5, TLC6 |
| Pat10  | 47,m  | SP,CP,GTC   | TLA1, TLB1, TLB2; TRB2, TRC2, TRC5 |
| Pat12  | 42,f  | SP,CP,GTC   | TBA4, TBB6, HR_7; TLB2, TLB3, TLC2 |
| Pat15  | 31,m  | SP,CP,GTC   | TBA1, TLR4, TLA4; HR_8, FRA6, FRB3 |
| Pat16  | 50,f  | SP,CP,GTC   | HL_2, HL_3, TBB1; G_H1, HL_9, TBC2 |
| Pat17  | 28,m  | SP,CP,GTC   | TBA1, TBA2, TLA1; G_B5, G_B8, G_C5 |
| Pat21  | 13,m  | SP,CP       | TBB1, G_E6, TBA3; G_A2, G_B2, G_C2 |

|        |       |             | *Temporo/Occipital* |
|--------|-------|-------------|---------------------|
| Pat06  | 31,f  | CP,GTC      | TLC1, TLC2, OBB1; TRC1, GD8, OBB4 |
| Pat09  | 44,m  | CP,GTC      | TBA3, TBB2, TBC4; G_A1, G_A8, TBC1 |
| Pat13  | 22,f  | SP,CP,GTC   | TBC1, TBC2, TBB1; POA1, POB1, TRA1 |

**Table I**   Brief information about the patients with focal epilepsies [32] grouped with regard to the origins of the epilepsies. The patients are designated on the left column (PatNM) and the numbers on the second column show the ages where f or m stands for female or male, respectively. The seizure types are shown on the third column which are simple partial (SP), complex partial (CP) or generalized tonic-clonic (GTC). The first three of the electrodes given on the fourth column are in-focus and the others (after the semicolons) are out-focus.

**FIGURES**

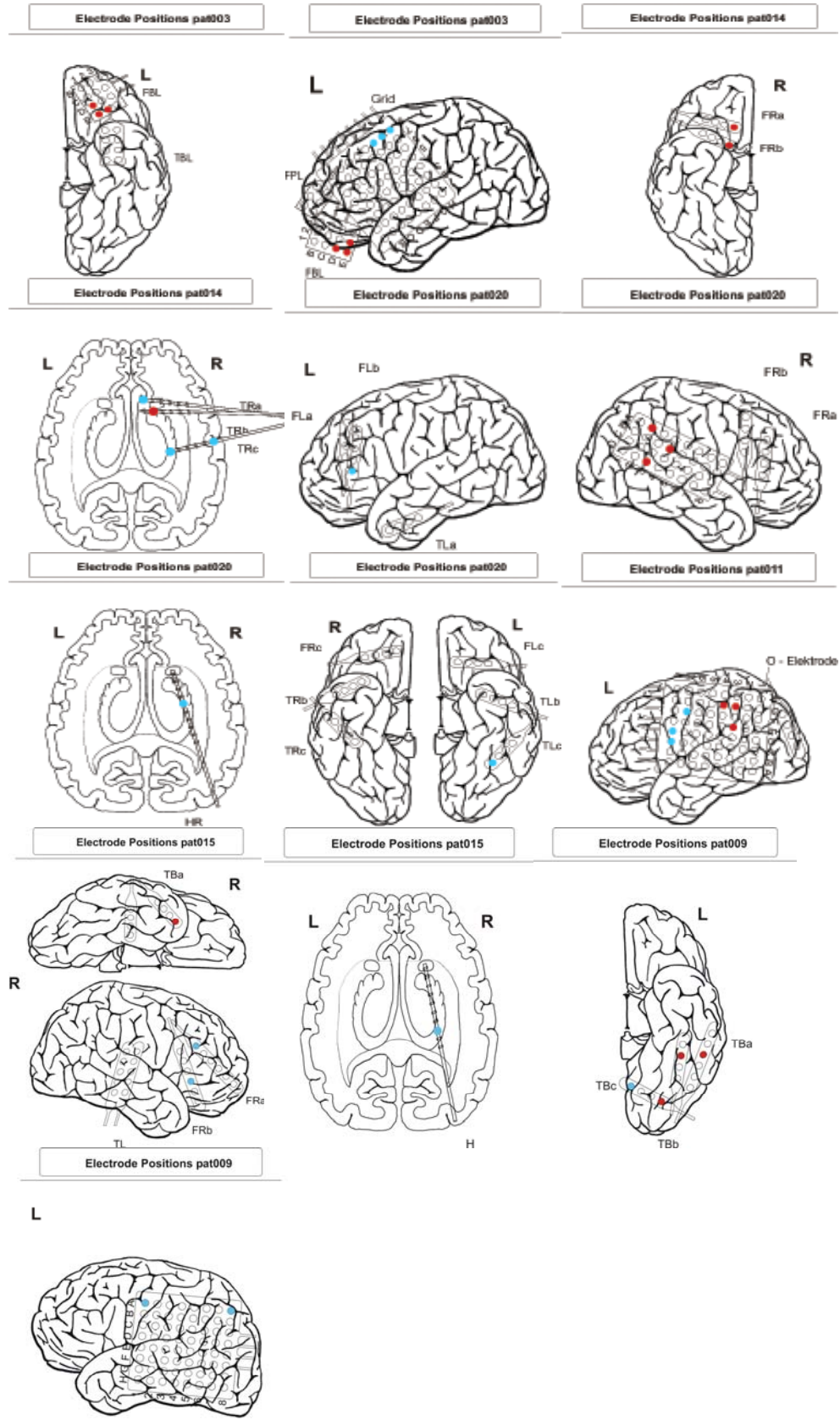

**Figure 1** The available electrode types and contact positions of the declared patients where the red and blue dots indicate in-focus or out-focus electrodes, respectively [32].

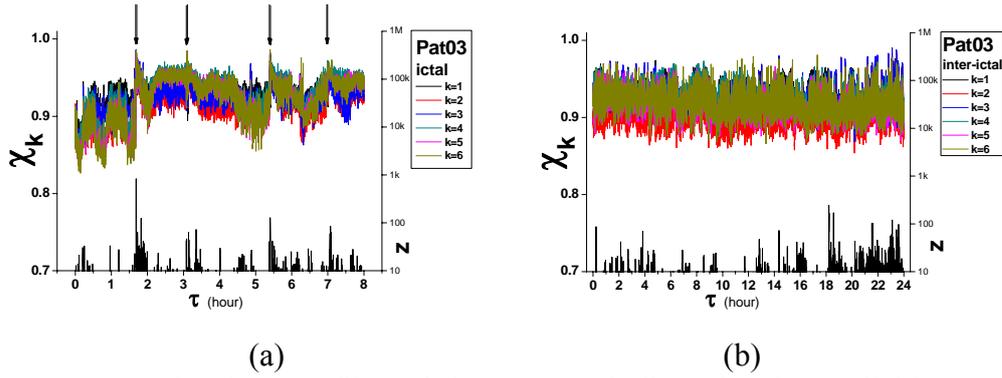

(a)  (b)

**Figure 2**   The time profiles of the entropy indices for the available [32] ictal (a) and inter-ictal (b) data from the Pat03 (see, Table I) where the arrows in (a) indicate the seizure terms. The black line at the bottom is for $z(\tau)$ in the plots.

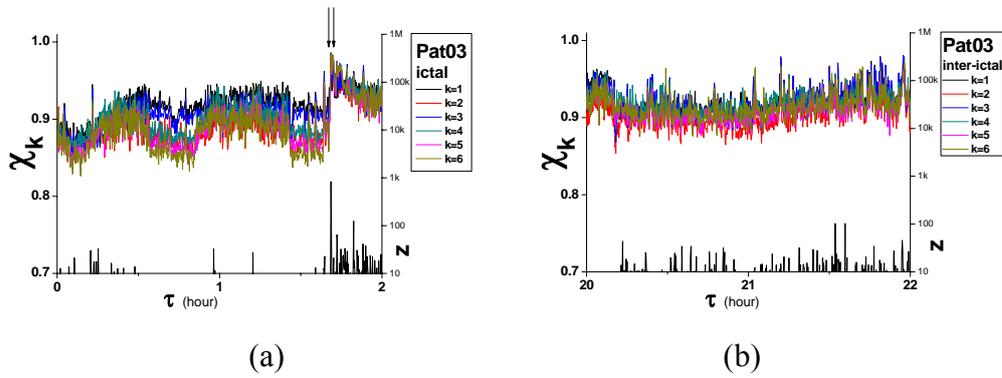

(a)  (b)

**Figure 3**   The plots (a) and (b) are same as Figs. 2 (a) and (b), respectively where the time domains are different.

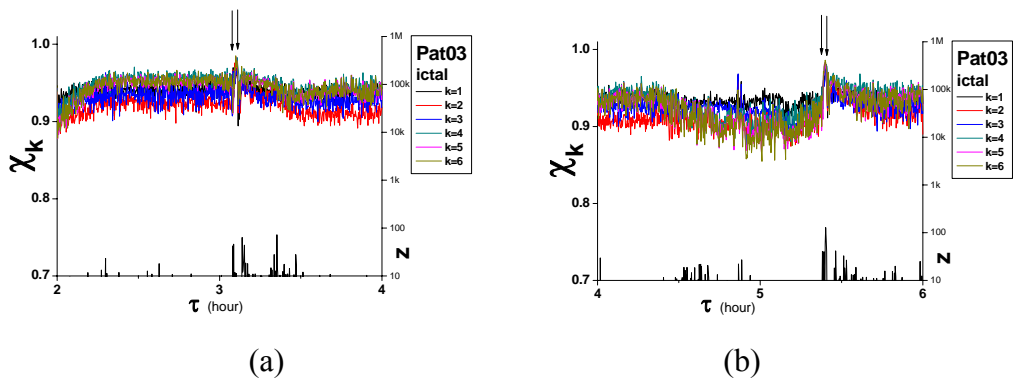

(a)  (b)

**Figure 4**   The plots are same as Figs. 2 (a) with different time domains.

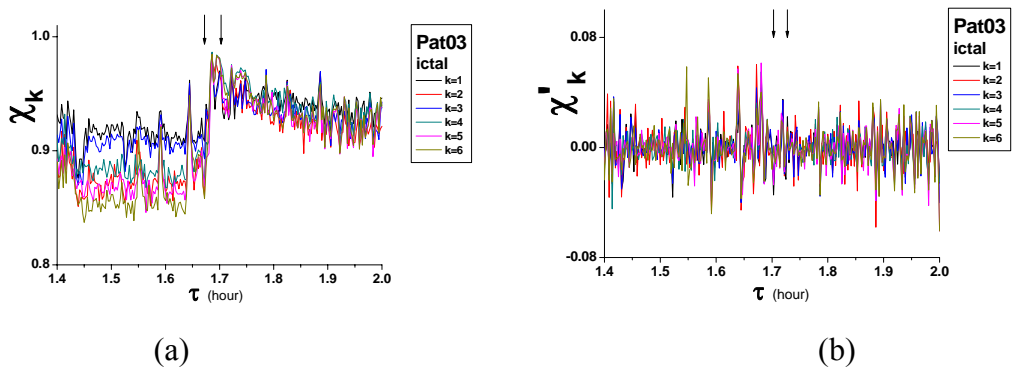

(a)  (b)

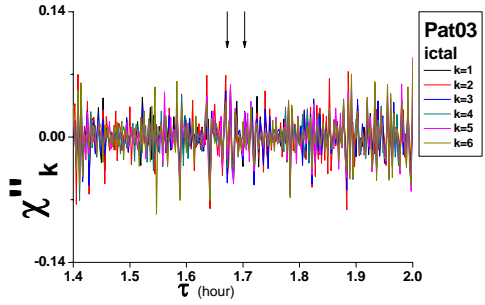

(c)

**Figure 5** The time series of $\chi_k(\tau)$ is depicted in (a) which is same as with 3 (a) for a shorter time domain. The successive time differences $\chi'_k(\tau)$ and $\chi''_k(\tau)$ are displayed in (b) and (c), respectively with the same time domain as in (a), here.

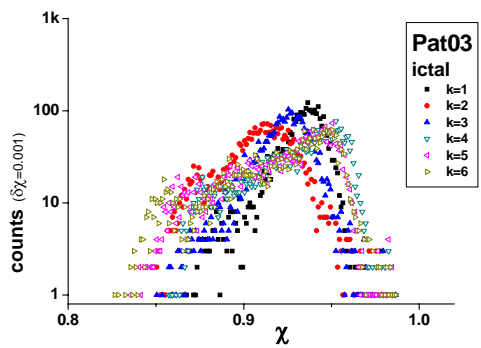
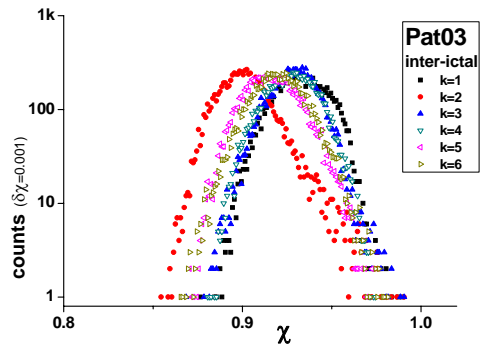

(a)                                           (b)

**Figure 6** The distributions of the entropy index values for all of the data from the patient in ictal interval (a) and in inter-ictal interval (b).

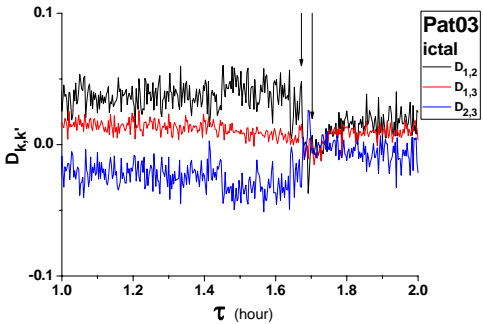
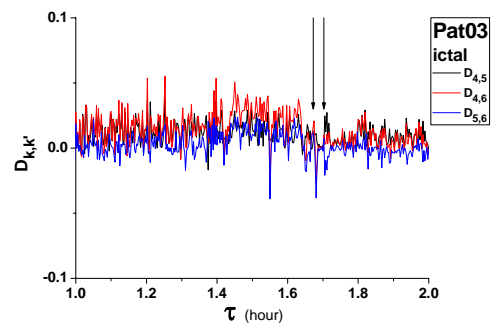

(a)                                           (b)

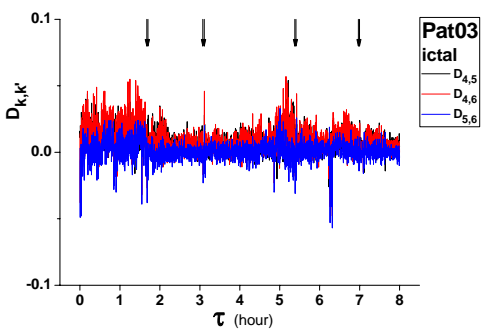

(c)

**Figure 7** The synchronous differences of the entropy index values of the in-focus electrodes (a) and out-focus electrodes (b) and (c) of the patient in ictal interval.

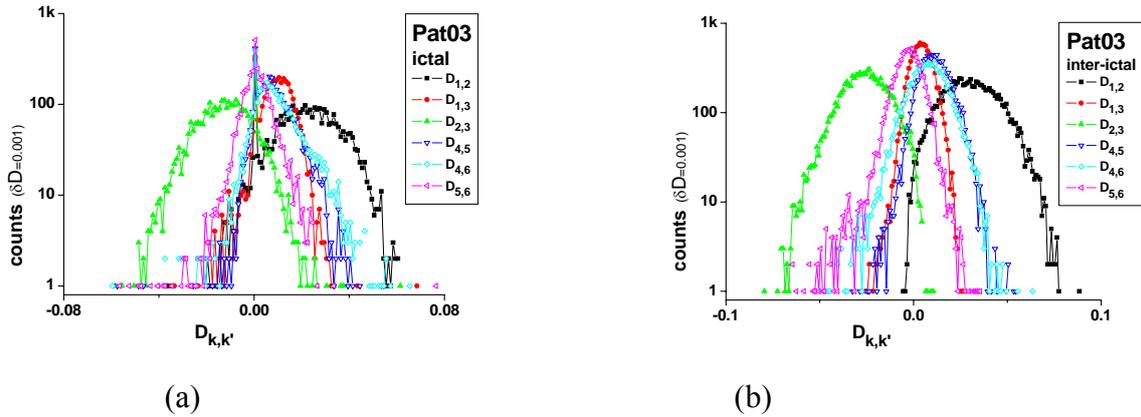

**Figure 8** The distribution of the values of the synchronous spatial differences of the in-focus electrodes of the Pat03 in ictal interval (a) and in inter-ictal interval (b).

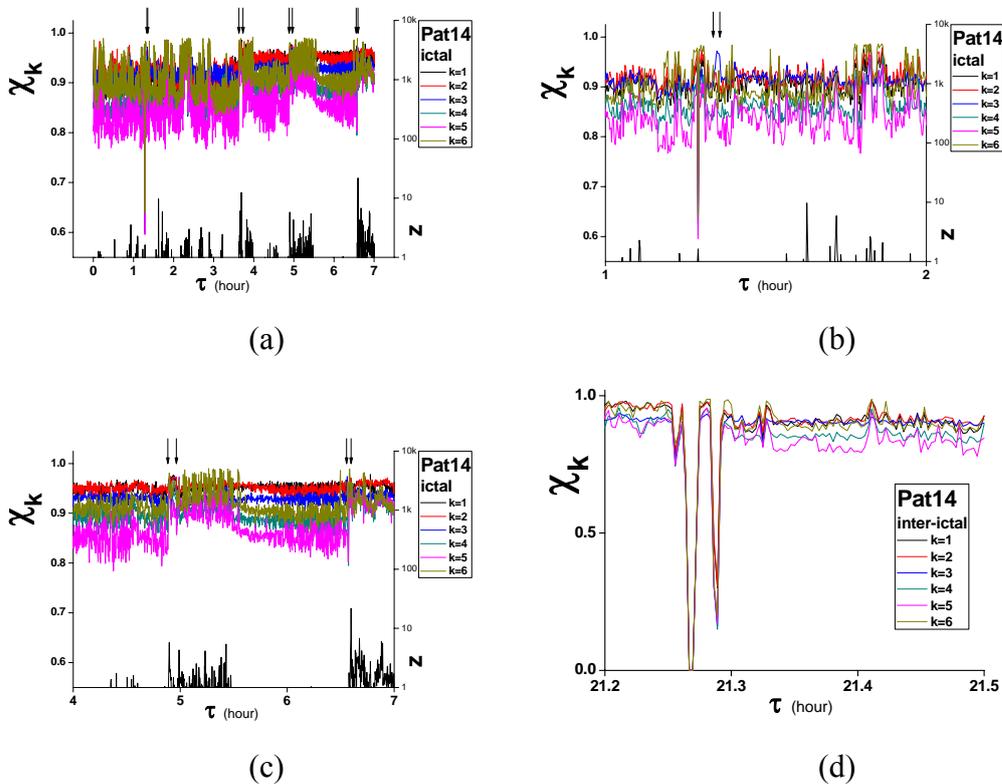

**Figure 9** Entropy index profiles from the Pat14 in ictal interval and in inter-ictal interval are shown in (a)-(c) and (d), respectively with different time domains.

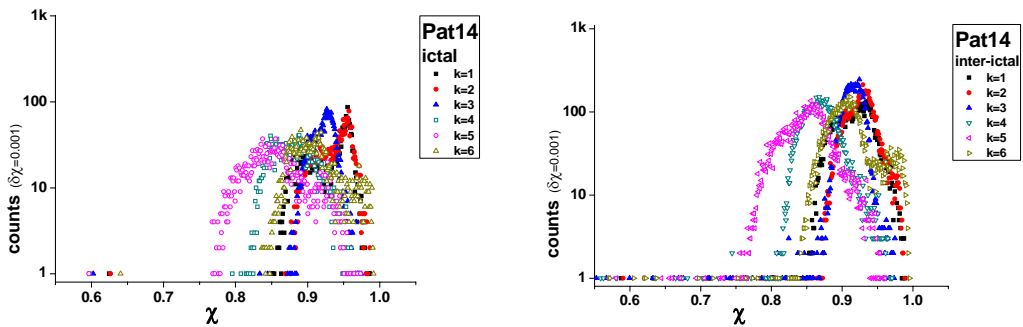

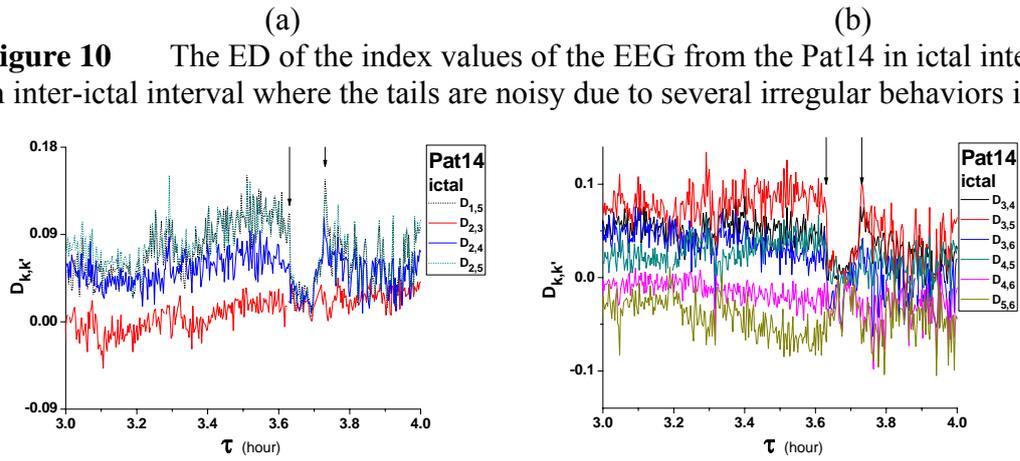

(a)                                        (b)

**Figure 10**      The ED of the index values of the EEG from the Pat14 in ictal interval (a) and in inter-ictal interval where the tails are noisy due to several irregular behaviors in the data.

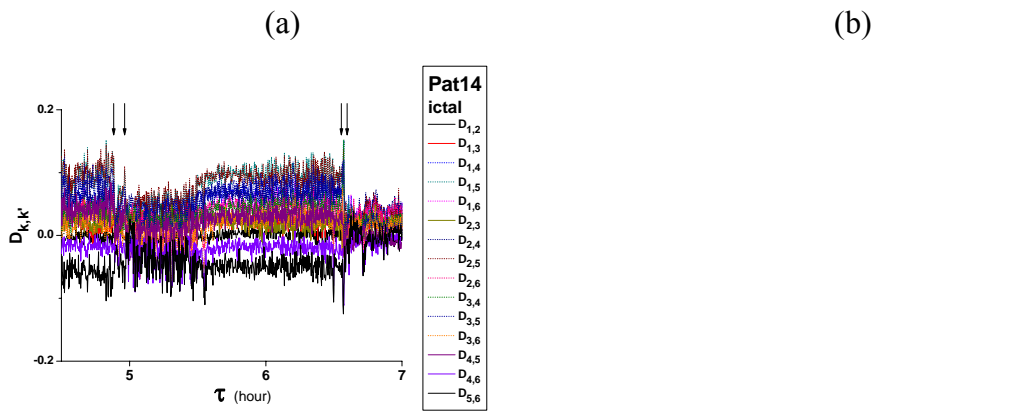

(a)                                        (b)

(c)

**Figure 11**      The time profiles of several local and non-local electrode couplings about the same (a) and (b) or different seizure terms (c).

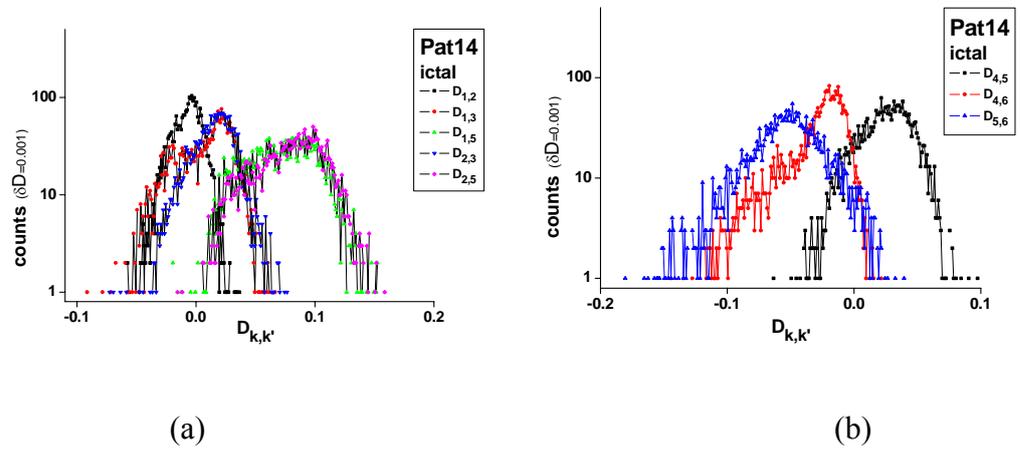

(a)                                        (b)

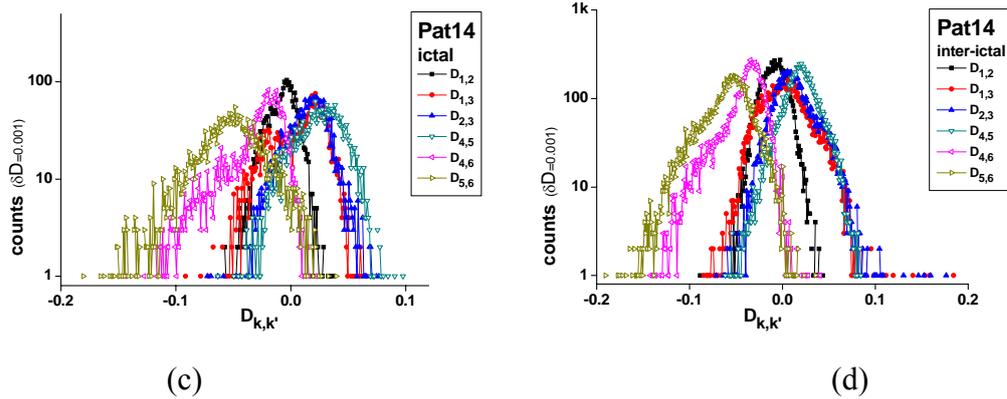

(c)                                                         (d)

**Figure 12**    Various cross-correlations between the in-focus and out-focus electrodes of the patient in ictal interval (a)-(c) and in inter-ictal interval (b).

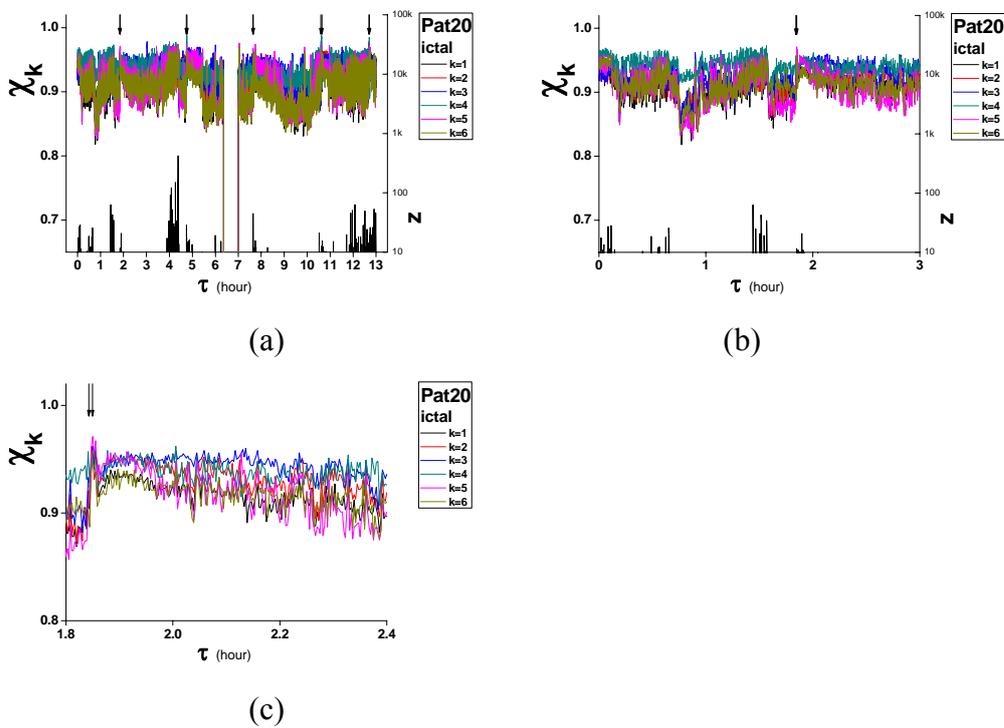

(a)                                                         (b)

(c)

**Figure 13**    The time profiles of the indices of the Pat20 in ictal interval with different time domains (a)-(c).

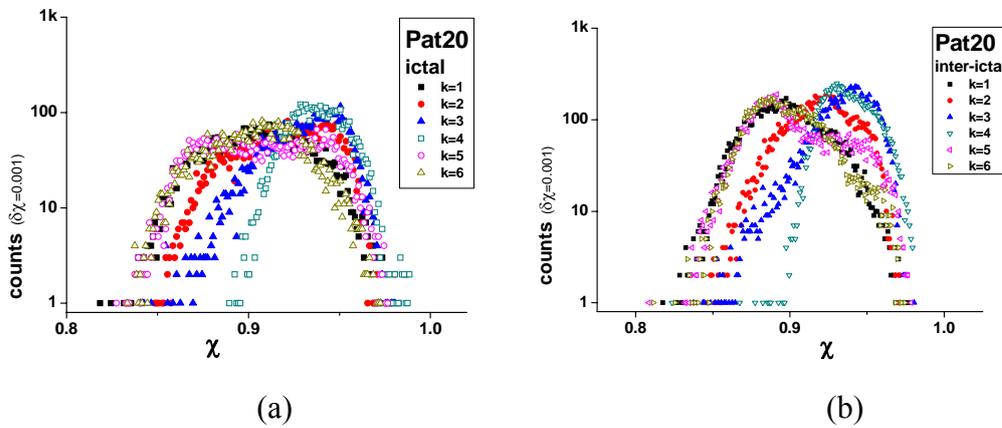

(a)                                                         (b)

**Figure 14**    The ED plots for the index values from the Pat20 in ictal interval (a) and in inter-ictal interval (b).

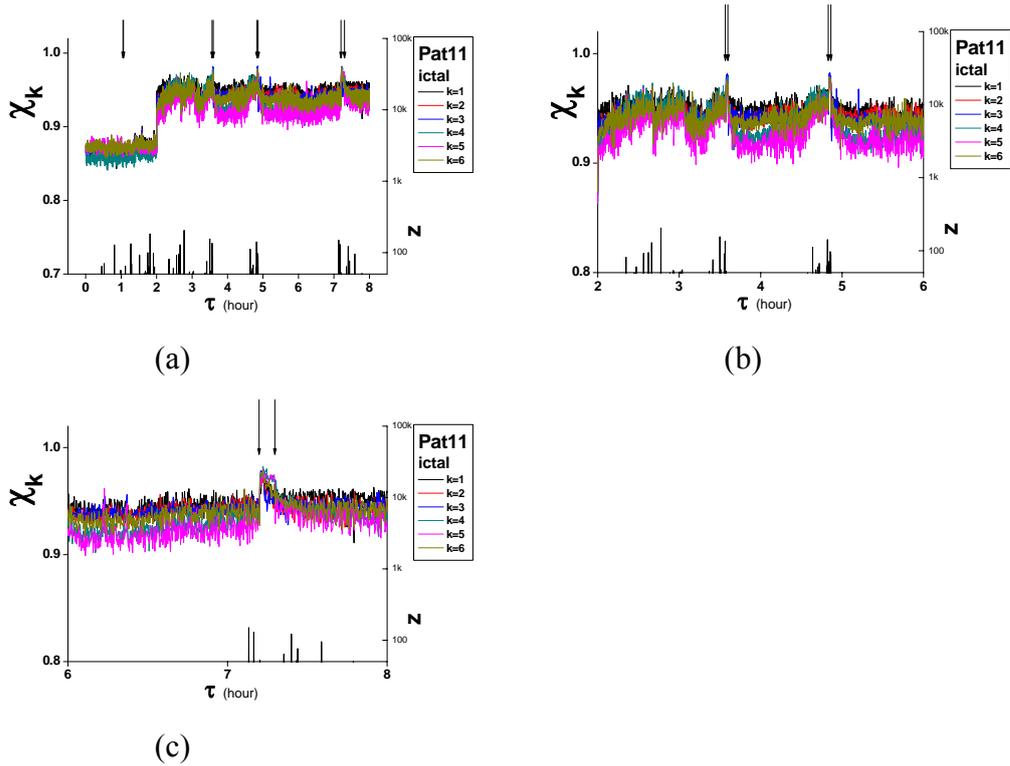

**Figure 15** The time profiles of the ictal entropy indices of the Pat11 in (a)-(c) with various time domains.

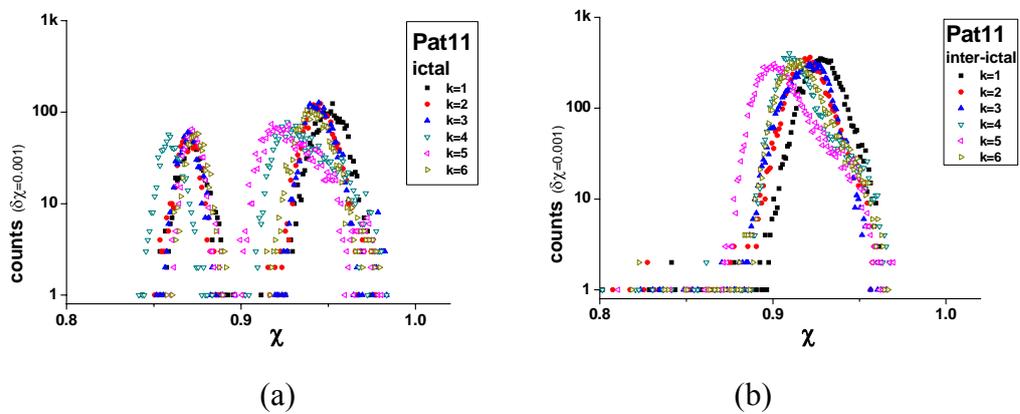

**Figure 16** The ED plots of the index values of the Pat11 in ictal interval (a) and in inter-ictal-interval (b).

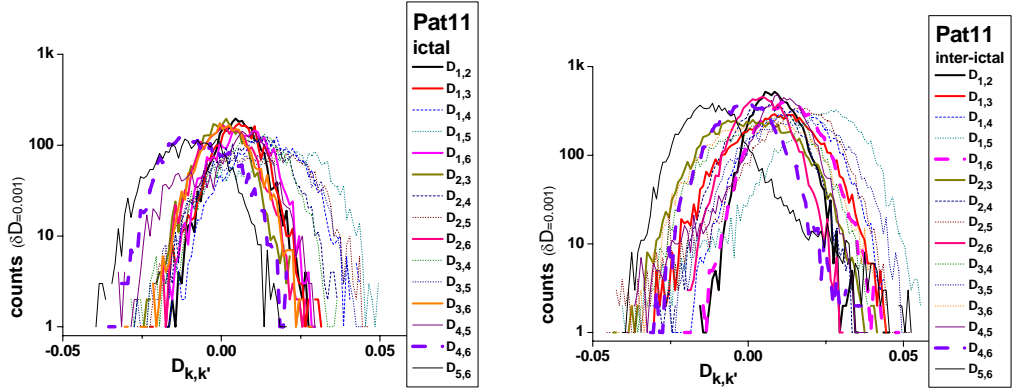

**Figure 17** The local and non-local cross-correlations between the electrodes of the Pat11 in ictal interval (a) and in inter-ictal interval (b).

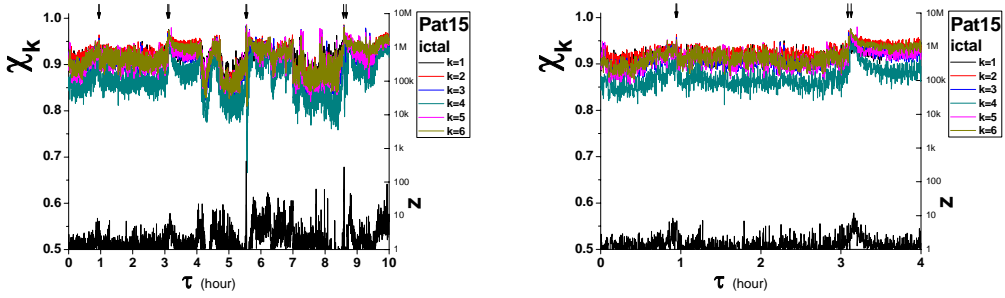

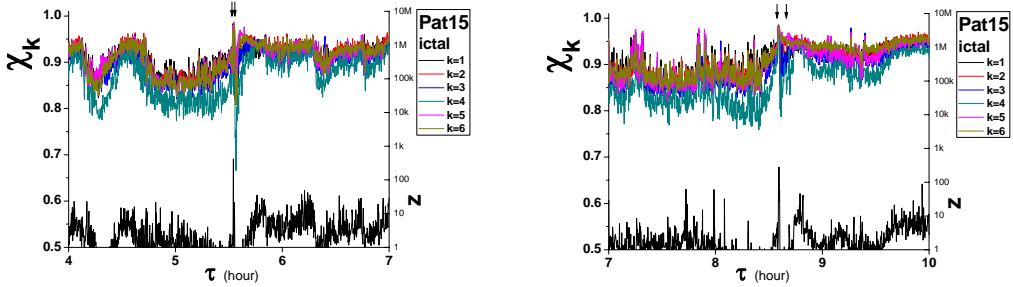

**Figure 18** The time profiles of the index values from the Pat15 in ictal interval with different time domains.

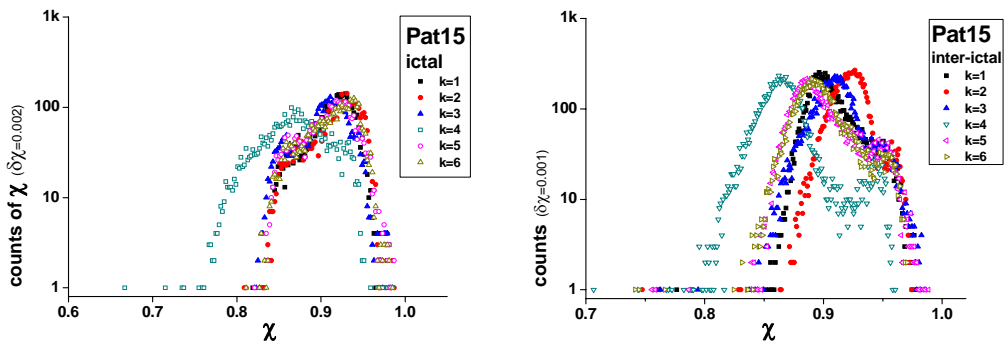

**Figure 19** The ED plots of the index values from the Pat15 in ictal interval (a) and in inter-ictal interval.

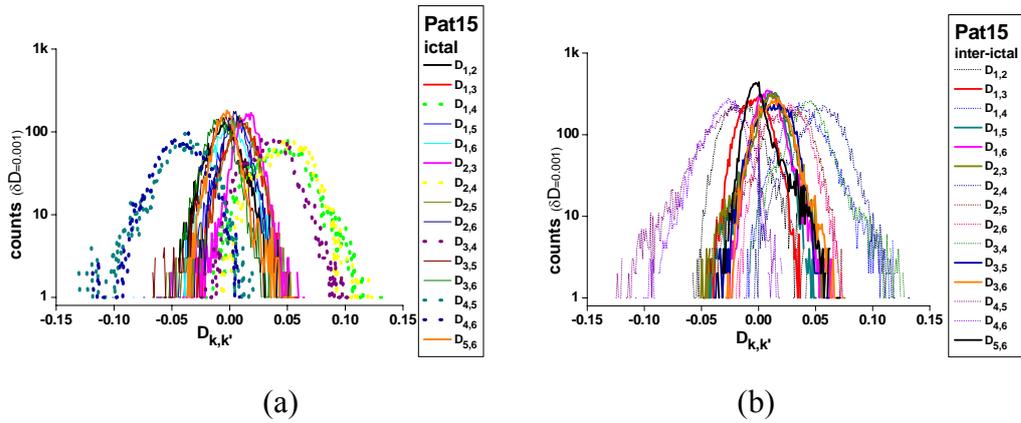

(a)          (b)

**Figure 20** The cross-correlations between the electrodes of the Pat15 in ictal interval (a) and in inter-ictal interval (b).

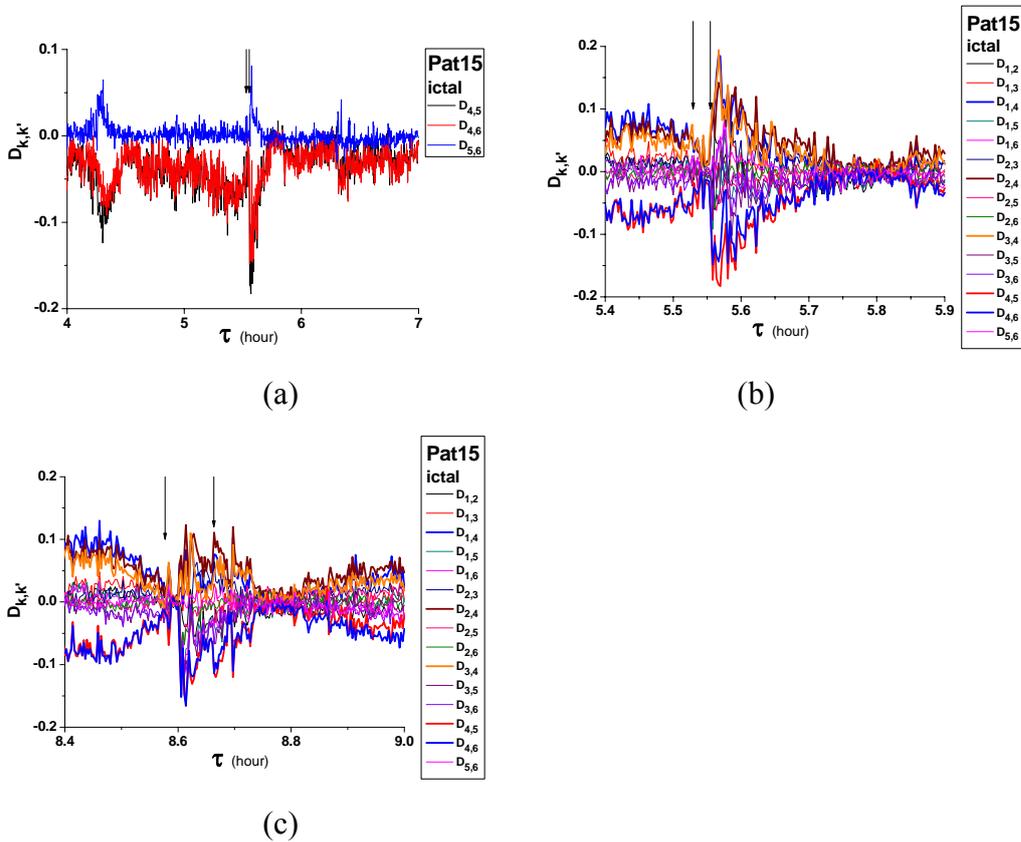

(a)          (b)

(c)

**Figure 21** The cross-correlations between the out-focus electrodes (a) and all of the electrodes ((b) and (c)) of the Pat15 in ictal interval within different time domains.

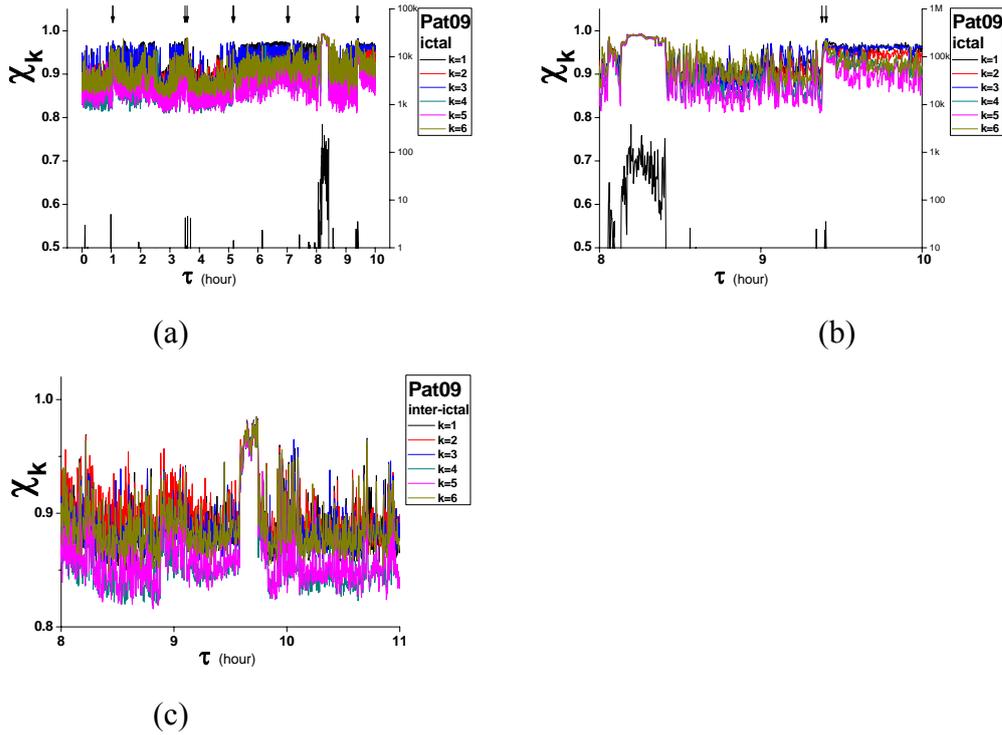

**Figure 22** The time profiles of the index values from the Pat09 in ictal interval (a) and (b) and in inter-ictal interval (c) with different time domains.

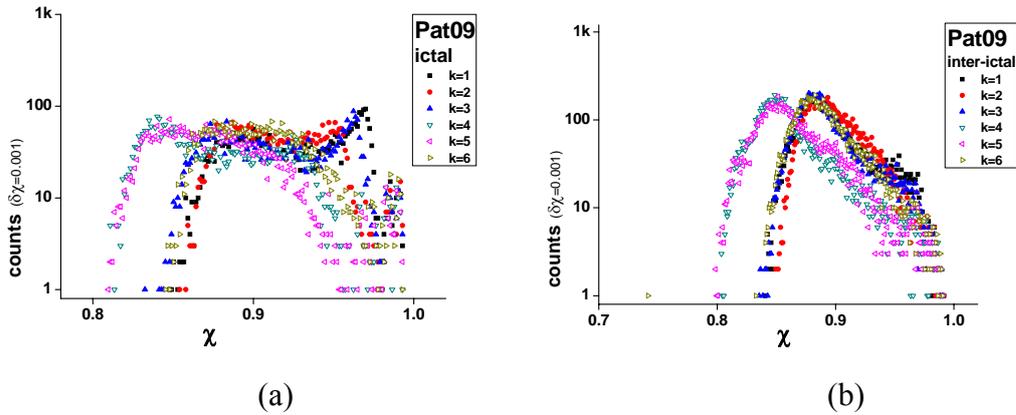

**Figure 23** ED plots of the index values from the Pat09 in ictal interval (a) and in inter-ictal interval (b).

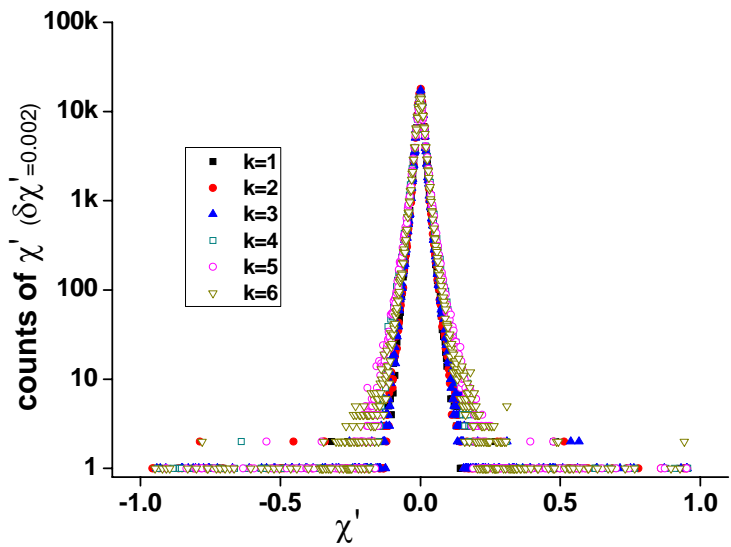

**Figure 24** The distributions of the values $\chi'_k(\tau)$ which are obtained using all of the data from 20 patients in ictal interval or in inter-ictal interval.

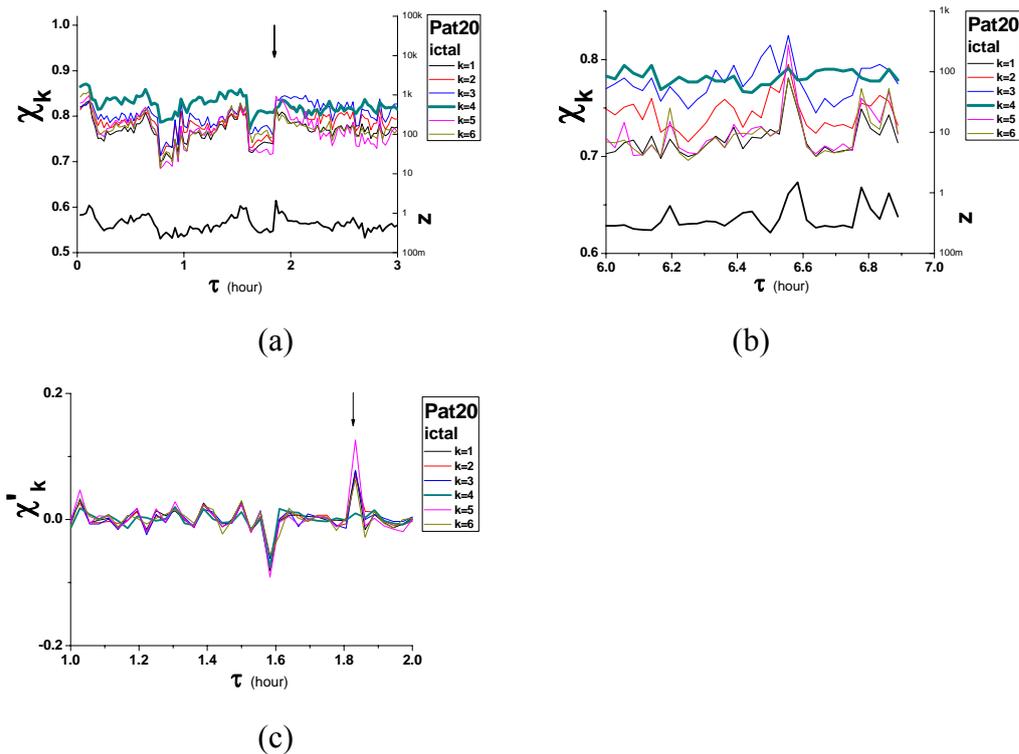

**Figure 25** Various applications of the method with a window time length of 100 seconds where (a) and (b) show the indices and (c) is for the temporal differences of the indices.